\documentstyle[12pt]{article}
\begin{document}
\renewcommand{\theequation}{\thesection.\arabic{equation}}
\vspace*{2cm}
\begin{center}
{\LARGE\bf Symmetry properties of the effective }
\end{center}
\begin{center}
{\LARGE\bf action for high-energy  scattering in QCD\footnote{ Collaboration
supported in part by Volkswagen Stiftung} }
\end{center}
\vspace{1cm}
\begin{center}
{\large\bf R. Kirschner}\\{\bf DESY-Institut f\"ur Hochenergiephysik }\\
{\bf Zeuthen, Germany  }
\end{center}
\begin{center}
{\large\bf L.N. Lipatov\footnote{Supported in part by the Russian
Foundation of Fundamental Investigations, grant No. 92-02-16809.}}\\{\bf
St.
Petersburg Nucl. Phys. Institute}\\ {\bf Gatchina, Russia} \end{center}
\begin{center}
{\large\bf L. Szymanowski\footnote{ Supported by the
Polish KBN Grant No. 2P302 143 06}}\\{\bf Soltan Inst. for Nuclear
Studies}\\{\bf Warsaw, Poland}
\end{center}
\vspace{1.5cm}

We  study the  effective action describing high-energy scattering
processes
in the multi-Regge limit of QCD, which should provide the starting point
for a new attempt to overcome the limitations of the leading logarithmic
and the eikonal approximations. The action can be obtained via simple
graphical rules or by integrating in the QCD functional integral over
momentum
modes of gluon and quark fields that do not appear explicitely as
scattering
or exchanged particles in the considered processes. The supersymmetry is
used
to obtain the terms in the action involving quarks fields from the pure
gluonic ones. We observe a Weizs\"acker - Williams type relations between
terms describing scattering and production of particles.

\newpage
\section{Introduction}
\hspace*{4mm} In experiments at collider energies kinematical regions become
accessible, that are characterized by two momentum scales, both much
larger than the typical hadronic scale $\mu $. Processes dominated by
this kinematics are called semi-hard \cite{LevRys}. The main features of
such processes are calculable perturbatively. Unlike in usual hard
processes, large contributions having typically a logarithm of the large
scales from each loop change the result qualitatively compared to the
Born contributions.

  If the largest scale determines the scattering energy and the second
scale determines the typical transverse momentum or the momentum
transfer, then the semi-hard kinematical region falls into the Regge
region. This situation we encounter in deep-inelastic scattering at
small values of the Bjorken variable $x$. Here we have for the
scattering energy squared $ s \approx -\mbox{Q}^2 / x $, where
$\mbox{Q}^\mu$  is the
momentum transferred by the photon, and therefore we have
\begin{equation}
        s \gg -\mbox{Q}^2 \gg \mu^2
\end{equation}
The low-$x$ region of structure functions becomes important in hard
processes at higher energies. Mini-jet production in hadronic
collisions provides a further example of semi-hard Regge processes.
Clearly here the separation of the perturbative contribution is more
difficult compared to the first example. Nevertheless at increasing
energies the semi-hard features should show up more and more clearly,
for example in selected final states with a large rapidity gap between
two jets, as proposed in \cite{Mueller}.

The perturbative Regge asymptotics has been investigated in the leading
logarithmic approximation (LLA)\cite{FKL}. The resulting amplitudes
do not satisfy
the s-channel unitarity constraints, and in particular a power-like
increasing contribution to the total cross section is obtained. In the
structure functions this appears as a fairly strong increase at small
$x$, i.e. the gluon density becomes so large that the concept of
quasi-free partons is not applicable any more.

   The eikonal approximation developed by Cheng and Wu \cite{ChengWu1}
for abelian gauge theory ensures s-channel unitarity
  (but it does not ensure unitarity in subenergy channels).
The generalization
to the non-abelian case encounters difficulties, which are understood
from the point of view of the LLA. At high energy the scattering is
dominated by the exchange of reggeized gluons with multi-particle
intermediate states in the s-channel (perturbative QCD pomeron).
In the abelian case the scattering amplitude with photon quantum
numbers in $t$-channel
gets no corrections in LLA.
In QED multi-particle s-channel states become important only on the
level of the $e^4 \ln s$ approximation ($e^{+} e^{-}$ pair production )
\cite{FrGr}.

Therefore there is no eikonal scheme relying only on the elastic Born
amplitude in the non-abelian case. An eikonal scheme starting from the
pomeron is used in some models. However this does not provide an
approximation to QCD Regge asymptotics, because the interactions of
the QCD pomerons are not negligible. A pomeron eikonal approach to the
problem of low-$x$ asymptotics of structure functions has been developed
in \cite{LevRys}.

High-energy scattering in the Regge regime has been studied in gravity
in order to understand quantum gravity and strings \cite{t'Hooft}
(in accordance with graviton reggeization).
 The eikonal scheme works here , but it neglects
 the multi-graviton intermediate states, which contribute in the LLA
 because of the graviton self-coupling \cite{LipGrav}. As in
non-ablian gauge theories the conventional eikonal scheme has to be
replaced by a unitarization scheme based on the LLA results \cite{AmCiVe}.

Including multi-particle production the unitarity conditions become more
involved. Unitarity has to be obeyed in all sub-energy channels of the
inelastic amplitudes \cite{StappWhite}, \cite{Bartels1}.
The dominating contribution in LLA comes from multi-particle states
obeying the conditions of multi-Regge kinematics (see Fig.~1)

\begin{eqnarray}
  s &=& (\mbox{p}_A + \mbox{p}_B)^2 = 2(\mbox{p}_A \mbox{p}_B) \nonumber \\
  s_i &=& (\mbox{k}_i + \mbox{k}_{i-1})^2 = 2(\mbox{k}_i \mbox{k}_{i-1})
          \qquad i = 1, \ldots , n+1 \nonumber \\
  \mbox{k}_0 &\equiv& \mbox{p}_{A'}, \qquad \mbox{k}_{n+1} \equiv
          \mbox{p}_{B'}, \qquad \mbox{k}_i = \mbox{q}_{i+1} - \mbox{q}_i
          \nonumber  \\
  s &\gg& s_1 \sim s_2 \sim \ldots \sim s_{n+1} \gg |\mbox{q}_1^2| \sim
       |\mbox{q}_2^2| \sim \ldots \sim |\mbox{q}_{n+1}^2| \nonumber \\
  s_1 s_2 &\ldots& s_{n+1} = s \prod^n_{i=1} (-k^2_{\perp i}) \nonumber
  \\
  &\mbox{k}_1\mbox{p}_A& \ll \mbox{k}_2\mbox{p}_A \ll \ldots \ll
  \mbox{k}_n\mbox{p}_A     \nonumber \\
  &\mbox{k}_1\mbox{p}_B& \gg \mbox{k}_2\mbox{p}_B \gg \ldots \gg
  \mbox{k}_n\mbox{p}_B
\end{eqnarray}
where $k_{\bot}^\mu$ is defined by the decomposition
\begin{equation}
\mbox{k}^\mu = \frac{(\mbox{p}_A \mbox{k})}{(\mbox{p}_A \mbox{p}_B)}
        \mbox{p}_B^\mu + \frac{(\mbox{p}_B \mbox{k})}{(\mbox{p}_A \mbox{p}_B)}
        \mbox{p}_A^\mu + k_{\bot}^\mu \nonumber
\end{equation}
In the LLA the elastic amplitude is dominated by the exchange of two
reggeized gluons in the colour singlet states interacting by the
exchange of gluons in s-channel
 \cite{FKL}. To obey s-channel unitarity contributions from the
exchange of more than two reggeized gluons have to be included.
These contributions are higher order corrections from the point of view
of the LLA. We keep the condition of the s-channel multi-particle states
being in multi-Regge kinematics. In this sense we would like
to include only a minimal
set of higher order corrections to the LLA. Contributions from two or
more particles being in non-multi-Regge kinematics are not included in
the first step. They lead to corrections to the scattering and
production vertices considered below and to new vertices. A program to
investigate such corrections has been started in \cite{FadLip}.

The equations for the contribution of the exchange of more than two
Reggeized gluons turn out to be fairly complicated
\cite{Bartels2}. Moreover we cannot
restrict ourselves to a small number of reggeized gluons. Also the
representation of the problem in form of a reggeon calculus is not very
helpful because of the complicated multi-reggeon interactions.
As a new idea the effective action approach was proposed \cite{Leff}.

The effective action describes the scattering and production of
particles
in the multi-Regge kinematics. In particular, the tree approximation of the
multiparticle amplitude reproduces the leading contribution from the sum of
tree  diagrams of the QCD. Compared to original QCD this formulation is
significantly simpler since it does not involve fields with Lorentz and
spinorial structure. This is related to the fact that in the LLA the helicities
of scattered particles are conserved.

In the considered multi-Regge kinematics there is a natural separation of the
longitudinal and transverse directions with respect to the scattering axis.
The structure of the effective lagrangian reflects this separation and in
particular one can  achieve its form emphasizing the scale and conformal
symmetries separately in the longitudinal and transverse subspaces. The
conformal symmetry in the impact space turned out to be a useful
tool for investigating the partial waves for the QCD pomeron and the odderon
\cite{LipKir}, \cite{Gln}.
This gives reason to expect that this effective action can be transformed
(in the impact space) to a two-dimensional model  exhibiting  full
conformal symmetry and will permit to apply the powerful methods developed
for  the two-dimensional conformal models.

The paper \cite{Leff} dealt with  pure gluodynamics and with the quantum
gravity.
 In a previous paper \cite{KLS} the effective action for full QCD
including quarks has been obtained directly from the original QCD action
by eliminating with the help of equations of motion certain field modes.
In the present paper we discuss in detail some aspects of the derivation
and symmetry properties of the effective action.

In the next section we derive from tree graphs the effective vertices
for scattering and production of quarks and gluons. By an appropriate
choice of the quark wave functions and the gluon polarization vectors
these vertices can be represented in a simple form. In order to write
down the effective action with these vertices one has to choose fields
describing the scattering and exchanged particles. Our choice of fields
is based on symmetry arguments discussed in the following sections.

We show that the effective action can  be obtained directly
from the original QCD action by separating the momentum modes of the
fields according to the multi-Regge kinematics and integrating out the
modes of highly virtual momenta. The fields appearing in the effective
action are now defined in terms of certain modes of the original gauge
and fermion fields.  In section 3 we outline the derivation
under a slightly different aspect compared to the paper \cite{KLS}.
We treat maily the gluon case and discuss shortly the generalization to
quarks. In section 4 we use supersymmetry arguments to obtain the
fermionic terms out of the gluonic ones.

 In section 5 we write the fermionic terms in such a way that the action
shows a clear separation of longitudinal and transverse subspaces.
Definite behaviour with respect to rotations and dilatations in both
subspaces can be attributed to the fields. We give the supersymmetry
transformation relating gluonic and fermionic terms.

Recently E. and H. Verlinde \cite{Verlinde} proposed an effective action for
high energy scattering in gauge theories based on elegant geometrical
 arguments. We comment on the relation to our multi-Regge effective
 action (see also \cite{KLS}).

\section{The effective vertices}
\setcounter{equation}{0}

\hspace*{4mm} In this chapter we derive the effective vertices and graphical
rules, which reproduce in a most economic way the leading contribution in
the multi-Regge kinematics (1.2) to multi-particle tree amplitudes. Then
the effective action is written down in such a way that it reproduces these
rules.

We start from tree-level helicity $2 \rightarrow 2$ and $2 \rightarrow 3$
amplitudes. It is essential to choose wave functions for gluon and quark
helicity states in a convenient way. The results for the tree amplitudes
allow to read off the graphical rules. For writing the effective action
we have to introduce fields, the asymptotic states of which correspond to
the helicity state wave functions.

 In what follows we decompose all the momenta according to Eq.(1.3). In the
similar way we also decompose the gluon wave function (for notational
simplicity the colour indices are suppressed)
 $\varepsilon^\mu(\mbox{k},\lambda)$ with the helicity \nolinebreak $\lambda$
\begin{equation}
\varepsilon^\mu(\mbox{k},\lambda) =
 \frac{(\mbox{p}_A \varepsilon)}{(\mbox{p}_A\mbox{p}_B)}\mbox{p}^\mu_B +
\frac{(\mbox{p}_B \varepsilon)}{(\mbox{p}_A\mbox{p}_B)}\mbox{p}^\mu_A +
\epsilon^\mu_\bot(\mbox{k},\lambda)
\end{equation}
Since in the multi-regge kinematics we have a separation of the longitudinal
and the transverse components it is very convenient to introduce the light-cone
variables and the complex transverse coordinates
\begin{eqnarray}
 k_{\pm} &=& \mbox{k}_0 \pm \mbox{k}_3 ,\quad
 k = k_{\bot}^1 + ik_{\bot}^2, \quad k^* = k_{\bot}^1 -ik_{\bot}^2
                              \; , \nonumber\\
\epsilon_{\pm}(\mbox{k},\lambda)
&=&\varepsilon_0(\mbox{k},\lambda)\pm \varepsilon_3(\mbox{k},\lambda),\quad
 \epsilon (\mbox{k},\lambda) = \epsilon_{\bot}^1(\mbox{k},\lambda) +
            i\epsilon_{\bot}^2(\mbox{k},\lambda) \; , \nonumber\\
\partial_{\pm} &=& \frac{1}{2}(\partial_0 \pm \partial_3),\quad \partial =
\frac{1}{2}(\partial_1 - i\partial_2), \quad \partial^* = (\partial)^*
 \; , \nonumber \\
\partial_- x_+ &=& \partial_+ x_- = \partial x =
\partial^* x^* = 1 \;\; .
\end{eqnarray}
We choose the incoming particles momenta as
 $\mbox{p}_A^\mu = \frac{\sqrt{s}}{2}(1,0,0,1)$ and
 $\mbox{p}_B^\mu = \frac{\sqrt{s}}{2}(1,0,0,-1)$.

The straightforward method to obtain the effective action uses
the explicit form of the helicity wave functions. In the case of
gluons these wave functions $\varepsilon^\mu (\mbox{k}, \lambda)$
can be written down in the axial gauge where the gauge fixing
vector is either the incoming momentum p$_A$ (L gauge) or the
incoming momentum p$_B$ (R gauge), as shown in Fig. 1. Moreover
they can be parametrized by their transverse components
$\epsilon^\mu_\bot (\mbox{k}, \lambda)$ (see \cite{Leff}). Then the
gauge transformation of the transverse components
$\epsilon^\mu_\bot (\mbox{k},
\lambda)$ of the gluon wave function
\begin{equation}
  \epsilon^\mu_{\bot L} (\mbox{k}, \lambda) = \epsilon^\mu_{\bot
    R} (\mbox{k}, \lambda) - 2 \frac{(k_\bot \epsilon_{\bot R})}{k^2_\bot}
    k^\mu_\bot
\end{equation}
written down in the complex number notation takes the form of
the phase transformation
\begin{equation}
  \epsilon_L (\mbox{k}, \lambda) = - \frac{k}{k^*} \left(
     \epsilon_R (\mbox{k}, \lambda)\right)^*
\end{equation}
(the subscripts $L,R$ denote the $L$ or $R$ gauges,
respectively). The gluon helicity states\footnote{The helicity
state in the axial gauge defined by the vector $n^\mu$
($n\mbox{k} \neq 0$) is an eigenvector of the rotation generator
belonging to the little group simultaneous of both vectors
k$^\mu$ and $n^\mu$.}
 have the forms \newline
a) in the $L$ gauge
\begin{eqnarray}
  \varepsilon_L^\mu  (\mbox{k},+)
    &=& e_L (\mbox{k}) \left\{ \frac{1}{\sqrt{2}}
   \left(\delta^\mu_1 - i \delta^\mu_2 \right) +
   \frac{k^*}{\sqrt{2s} k_+} \mbox{p}^\mu_A \right\} \nonumber \\
 \varepsilon_L^\mu (\mbox{k}, -) &=& \{ \varepsilon_L^\mu
    (\mbox{k}, +) \}^*
\end{eqnarray}
b) in the $R$ gauge
\begin{eqnarray}
  \varepsilon_R^\mu (\mbox{k}, -) &=& e_R (\mbox{k}) \left\{
     \frac{1}{\sqrt{2}} (\delta^\mu_1 - i \delta^\mu_2) +
     \frac{k^*}{\sqrt{2s} k_-} \mbox{p}^\mu_B \right\} \nonumber \\
  \varepsilon^\mu_R (k,+) &=& \{ \varepsilon^\mu_R (k,-) \}^*
  \end{eqnarray}
The helicity states given by eqs. (2.5),(2.6) are determined up to the
phase factors $e_L (\mbox{k})$ and $e_R (\mbox{k})$. As a
consequence of eq. (2.4) they are subject to the gauge transformation
\begin{equation}
  e_L (\mbox{k}) = - \frac{k}{k^*} (e_R (\mbox {k}))^*
\end{equation}

We proceed in the similar way in the case of quark fields. The
quark helicity states $u(\mbox{k}, \lambda)$ have the forms (again
the colour indices are suppressed)
 \newline
a) in the $L$ gauge
\begin{eqnarray}
  u_L(\mbox{k},+) &=& \frac{\chi_L(\mbox{k})}{\sqrt{2k_+}} \left(
    \begin{array}{l}
      \phi_L (\mbox{k},+) \\ \phi_L (\mbox{k},+) \end{array} \right)
  \;  ,  \quad \phi_L (\mbox{k},+) = \left( \begin{array}{l}
       k^* \\ k_+ \end{array} \right) \; ,  \nonumber \\
  u_L (\mbox{k}, -) &=& C[\overline{u}_L (\mbox{k},+)]^T \; ,
\end{eqnarray}
\hspace*{0.5cm} ($C$ is the charge conjugation matrix) \newline
b) in the $R$ gauge
\begin{eqnarray}
 u_R (\mbox{k},-) &=& \frac{\chi_R(\mbox{k})}{\sqrt{2k_-}}
   \left( \begin{array}{r}
   \phi_R (\mbox{k},-)  \\ - \phi_R (\mbox{k},-) \end{array}
   \right)\; , \quad \phi_R (\mbox{k},-) = - \left( \begin{array}{l}
   -k^* \\ k_- \end{array} \right) \; ,\nonumber \\
 u_R(\mbox{k}, +) &=& C [\overline{u}_R (\mbox{k},-)]^T \; .
\end{eqnarray}
In Eqs. (2.8),(2.9) the functions $\chi_L (\mbox{k})$ and $\chi_R
(\mbox{k})$ are arbitrary phase functions modulo which the
helicity states are determined (compare Eqs. (2.5),(2.6)). Since the
helicity wave functions $u_L$ and $u_R$ are physically
equivalent (i.e. $u_L (\mbox{k}, \lambda) = u_R (\mbox{k},
\lambda)$) we get
\begin{equation}
  \chi_L (\mbox{k}) =  \sqrt{\frac{k}{k^*}} \left[ \chi_R
     (\mbox{k}) \right]^*
\end{equation}
which is the relation analogous to the gauge condition (2.7).

Consider now as an example the tree approximation of the gluon production
in the quark gluon scattering. For  definitness we take the case that
all particles have the same helicities $\lambda = +$. As
known from  Refs. \cite{FKL} and \cite{FS} the result, being a
sum of many Feynman graphs, can be written in the factorized
form as shown in Fig.~2 (where gluons and fermions are denoted by
wavy and solid lines, respectively).
 Using the form of helicity states
introduced above and the complex number notation we obtain
\begin{eqnarray}
\lefteqn{
  \left[ gt^A_{A'n_1} e_R (\mbox{p}_A) \chi_R (\mbox{p}_{A'})
  \sqrt{2p_{A'-}}
  \right] \; \frac{1}{q^*_1}}
   \nonumber \\
 &&  \left[  g\sqrt{2}t^k_{n_1n_2} q^*_1
    \frac{e_L (\mbox{k})^*}{k^*} \right]
   \frac{1}{q^*_2} \;
  \left[ g t^{B'}_{n_2B} \sqrt{2p_{B+}} \chi_L (\mbox{p}_B) e_L
  (\mbox{p}_{B'})^* \right]  \; ,
\end{eqnarray}
where the matrices $t^a_{bc}$ are the colour group generators of
the quark representation. The expression (2.11) is a product of
the effective vertices corresponding to the scattering processes
appearing on both ends of Fig.~2 and the gluon production
vertex. They are connected by the propagators of the helicity
$+$ fermionic particles $\left( \frac{1}{q^*} \right)$. \\
We present also for further reference the expression describing
the helicity $-$ gluon production in two fermion scattering
(both having the helicities $+$). The tree scattering amplitude
(see Fig.~3) has the form
\begin{eqnarray}
& \left[ g \sqrt{2} t^{n_1}_{A' A}
\chi_R(\mbox{p}_{A'})\chi_R(\mbox{p}_A)^* \frac{1}{2}(p_{A -} +
p_{A' -}) \right] \; \frac{1}{q_1 q_1^*} \cdot
\left[ ig\sqrt{2}f^{k n_1 n_2} q_1 q_2^*
{{e_R(\mbox{k})^*}\over {k^*}} \right] & \; \cdot \nonumber \\
& \cdot \frac{1}{q_2 q_2^*} \;
\left[ g\sqrt{2} t^{n_2}_{B' B}
\chi_L(\mbox{p}_{B'})^*\chi_L(\mbox{p}_B) \frac{1}{2} (p_{B +} +
p_{B' +}) \right] &
\end{eqnarray}
where $f^{abc}$ are the colour group structure constants.In the
middle square bracket there appear the effective vertex
describing the gluon production from the $t$-channel gluon line
\cite{Leff}.

Another example is provided by the helicity $+$ fermion
production in the quark gluon scattering (again all gluons have
the helicities $+$) as is shown in Fig.~4. The tree
approximation scattering amplitude has the form
\begin{eqnarray}
\lefteqn{\left[ -i g\sqrt{2} f^{n_2A'A} \; e_R
   (\mbox{p}_A)^* \; e_R (\mbox{p}_{A'}) \frac{1}{2} (p_{A-} +
   p_{A'-}) \right] \; \frac{1}{q_1 q_1^*} \cdot}  \nonumber \\
   & & \cdot \left[ -g \sqrt{2} t^{n_1}_{kn_2} q_1^* \sqrt{k_+}
     \frac{\chi^*_L(\mbox{k})}{k^*} \right] \, \frac{1}{q^*_2} \,
     \left[ gt^{B'}_{n_2B} \sqrt{2p_{B+}} \chi_L (\mbox{p}_B) e_L
     (\mbox{p}_{B'})^* \right]
\end{eqnarray}
Again from Eq.~(2.13) one can read off the corresponding effective
vertices and the propagators. \\  As these examples show,
despite of the colour indices there are no any other indices.
This is an essential simplification in comparison with the
original  formulation presented in Refs.~\cite{FKL} and \cite{FS}.
Table 1 collects Feynman rules from which one can write down the
corresponding expression for arbitrary process under consideration.

We want to write the effective action which reproduces the above
tree formulas. For this the wave functions $e_L, e_R,
\chi_L, \chi_R$ have to be related to the wave functions of
fields appearing in the effective action. These relations are
\begin{eqnarray}
 \phi^* (\mbox{k}) = 2\sqrt{2}\frac{e_L (\mbox{k})}{k} \; ,&&
 \phi(\mbox{k}) = - 2\sqrt{2}\frac{e_R (\mbox{k})}{k} \; ,\nonumber \\
   \chi_+(\mbox{k}) =  \sqrt{2k_+} \frac{\chi^*_L
     (\mbox{k})}{k^*} \; , &&
     \chi_- (\mbox{k}) =  \sqrt{2k_-} \frac{\chi^*_R
(\mbox{k})}{k^*}  \; , \nonumber \\
      \chi^*_+(\mbox{k}) = \sqrt{2k_+} \frac{\chi_L
     (\mbox{k})}{k} \; , &&
     \chi^*_-(\mbox{k}) = \sqrt{2k_-} \frac{\chi_R
(\mbox{k})}{k}  \;   .
\end{eqnarray}
The Coulomb interaction in $t$-channel is carried by real $
A_{\pm}$ fields in the case of gluon
  and in the case of fermion
by complex $a_{\pm}$, $a^*_{\pm}$ fields \cite{KLS}. The gluonic
fields $A_\pm$ should not be identified with the light-cone
components of gluonic four-potential (see ch.3).

The  effective action reproducing the effective vertices and propagators
discussed above is given by
\begin{equation}
  S = S_k + S_{s-} + S_{s+} + S_p
\end{equation}
where
\begin{eqnarray}
  S_k &=& S_{ks} + S_{kp}   \nonumber \\
  S_{ks} &=& \int d^4 x \left[ - \frac{1}{2} (\partial^* \phi^a)
       \Box (\partial \phi^{a*}) + i \bar{\chi}^{\alpha *}_+
       \Box \partial \chi^{\alpha}_- + i \bar{\chi}^{\alpha}_+
       \Box \partial^*
      \chi^{\alpha*}_-  \right]   \nonumber \\
  S_{kp} &=& \int d^4 x \left[ - 2 A^a_+ \partial\partial^*
       A^a_- - i\bar{a}^{\alpha}_+ \partial^* a^{\alpha*}_-
       - i \bar{a}^{\alpha *}_+ \partial a^{\alpha}_-
        \; \right.  \nonumber  \\
        && \left. - i \bar{a}^{\alpha *}_-\partial a^{\alpha}_+
        - i \bar{a}^{\alpha}_-\partial^* a^{\alpha *}_+ \;
         \right]
\end{eqnarray}
\begin{eqnarray}
  S_{s-} &=& - g \int d^4 x \left[ \frac{i}{2} (\partial_-
          \partial^* \phi) T^a (\partial \phi^*) A^{a}_+
     - \frac{i}{2}(\partial^*\phi)T^a(\partial_-\partial\phi^*)
     A^a_+ \right. \nonumber \\
     && \left. - [\bar{\chi}^*_+ t^a
      (\stackrel{\leftrightarrow}{\partial}_-\partial\chi_-)]
    A^{a}_+
    - [\bar{\chi}_+t^a
    (\stackrel{\leftrightarrow}{\partial}_- \partial^* \chi^*_-)]
    A^a_+ \right. \nonumber \\
 &&  \left. -  \bar{a}^*_+t^a(\partial_-\chi_+)
      (\partial\phi^{a *})  - \bar{a}_+t^a(\partial_-\chi^*_+)
      (\partial^*\phi^a)  \right. \nonumber  \\
 &&  \left. - (\partial_-\bar{\chi}^*_+)t^aa_+(\partial\phi^{a *})
       - (\partial_-\bar{\chi}_+ )t^aa^*_+(\partial^*\phi^a)
      \; \right]   \\
   S_{s+} &=& - g \int d^4 x \left[ \frac{i}{2} (\partial_+
     \partial^* \phi^*) T^a (\partial \phi) A^{a}_-  -
  \frac{i}{2}(\partial^*\phi^*)T^a(\partial_+\partial\phi)
     A^a_- \right. \nonumber \\
     && \left. - [\bar{\chi}^*_- t^a
     (\stackrel{\leftrightarrow}{\partial}_+\partial\chi_+)]
    A^{a}_-
    - [\bar{\chi}_- t^a
    (\stackrel{\leftrightarrow}{\partial}_+ \partial^* \chi^*_+)]
    A^a_- \right. \nonumber \\
  && \left. - \bar{a}^*_-t^a(\partial_+\chi_-)(\partial\phi^a)
     - \bar{a}_-t^a(\partial_+\chi^*_-)(\partial^*\phi^{a *})
     \right.  \nonumber \\
  &&  \left. - (\partial_+\bar{\chi}^* _-)t^aa_-( \partial\phi^
    a) - (\partial_+\bar{\chi}_-) t^aa^*_-(\partial^* \phi^{a
     * })  \; \right]    \\
  S_p &=& g \int d^4 x \left[ \phi^a(\partial A_-)T^a(\partial^*
      A_+) - \phi^{a *}(\partial^*
     A_-)T^a(\partial A_+ )   \right. \nonumber  \\
      &-& \left. \frac{i}{2}[\phi^a(-\bar{a}^*_+t^a(\partial
       a _-) + (\partial^*\bar{a}_+)t^aa^*_-)    \right.
     \nonumber  \\
     &+& \left. \phi^{a *}(\bar{a}_+t^a(\partial^* a^*_-) -
      (\partial\bar{a}^*_+)t^aa_-)  \right. \nonumber \\
    &+& \left. \phi^a(-(\partial\bar{a}^*_-)t^aa_+ +
    \bar{a}_-t^a(\partial^*a^*_+))   \right.  \nonumber  \\
    &+& \left. \phi^{a *}((\partial^*\bar{a}_-)t^aa^*_+ -
    \bar{a}^*_-t^a(\partial a_+))]  \right. \nonumber \\
    &+& \left. i[\bar{\chi}_-t^aa^*_-(\partial^* A^a_+) -
    \bar{\chi}^*_-t^aa_-(\partial A^a_+) \right.
     \nonumber \\
   &+&  \left. \bar{a}_-t^a\chi^*_-(\partial^* A^a_+) -
        \bar{a}^*_-t^a\chi_-(\partial A^a_+) \right. \nonumber \\
   &+& \left. \bar{a}_+t^a\chi^*_+(\partial^* A^a_-) -
      \bar{a}^*_+t^a\chi_+(\partial A^a_-) \right.
       \nonumber   \\
    &+& \left. \bar{\chi}_+t^aa^*_+ ( \partial^* A^a_-) -
       \bar{\chi}^*_+t^aa_+(\partial A^a_-]
             \right]
\end{eqnarray}

The bar over fermionic fields denotes the complex conjugation (
only for the Majorana particles we have $\bar{\chi} = \chi^*$,
$\bar{a} = a^*$).
The indices $a$ and $\alpha$ refer to the adjoint representation of the gauge
group  and the quark representation,  respectively.
In brackets bilinear in gluon fields the trace in the adjoint representation
is understood. $T^a$ are the matrices representing the
generators of the adjoint representation
\[ ({T^a})_{bc} = - if^{abc}, \;\;\;\; ({\cal A}_1 T^a
{\cal A}_2) = - if^{abc}{\cal A}^b_1 {\cal A}^c_2 \]
In brackets bilinear in the quark fields  the sum
over the gauge group indices of the quark representation  $t^a$
and over flavours is understood. The operator
 $ \Box = 4 (\partial_+ \partial_- - \partial\partial^*),
\stackrel{\leftrightarrow}{\partial}~=~\stackrel{\rightarrow}
{\partial}~-~\stackrel{\leftarrow}{\partial}$.

 In the next
chapters we present the symmetry arguments which justify the
choice of fields given by Eq.(2.14) and lead to the effective
action (2.15).

\section{Effective action from the QCD functional integral}
\setcounter{equation}{0}

\hspace*{4mm} In the previous chapter the effective action was obtained by
analysing the leading contributions of tree Feynman graphs. On
the other hand  effective action is conventionally
understood as a result of some fields appearing in the path
integral representation  being
integrated out.The aim of the present chapter is to show that
also the effective action (2.15) can be obtained in such a
way i.e. by integrating over certain modes  in the original
QCD action. This integration procedure differs in some aspect
from those which one encounters most frequently. In the usual
situation one integrates over all modes of the fields which do
not appear in the initial and final states.
In our case, due to the multi-Regge kinematics (1.2), the
outgoing particles are close to one of the incoming momenta
${\mbox{p}_A}$ or ${\mbox{p}_B}$. The virtual particles transfering
the momentum through $t$-channel carry small longitudinal
momenta. Because of this separation the integration over modes
which are absent in the resulting effective action is more
sophisticated than in the more familiar cases (Heisenberg-
Euler electrodynamics or the integration over heavy fields).
 A similar asymmetric separation of modes
appears in the derivation of the effective action for high
temperature limit.

We consider the high-energy scattering of gluons and quarks of
peripheral type, i.e. with momentum transfer of order
$\mu_\bot$, where $s \gg \mu^2_\bot \gg \mu^2$. In the leading
contribution essentially all produced particles are in momentum
space either close to the incoming particle p$_A$ or close to
p$_B$.

Let us start the discussion with the pure gluonic case. We shall
concentrate first on the particles flying approximately in the
direction of the incoming momentum p$_A $ and treat the other incoming
gluon as an external source.  In the light-cone axial gauge
 $A_{-}^a =0 $ the gluonic part of the QCD action expressed in terms of
the light-cone components $A_{\pm}^a$ and the transverse components
$A^{a \sigma}_{\perp}\equiv A^{a \sigma}$ (Eq.(2.1)) takes the form
\begin{eqnarray}
S_{g}&=& \int d^4x \lbrace {1 \over 2 } A_{\sigma }^{a} \Box A^{a \sigma } +
{1 \over 2 } (\partial_{-} A_{+}^{a} + \partial_{\sigma } A^{a \sigma
 })^2
- g f^{abc} (A_{\sigma }^{a} {{\partial}_-} A^{b \sigma }) A_{+}^{c}
\nonumber \\
&-&g f^{a b c }A_{\sigma }^{b} A_{\rho }^{c} \partial^{\sigma }A^{a \rho
} - {g^2 \over 4} f^{a b c } f^{a b^{\prime } c^{\prime } }
A_{\sigma }^{b} A_{\rho }^{c} A^{b^{\prime } \sigma } A^{
c^{\prime } \rho} \rbrace
\end{eqnarray}
We define
\begin{equation}
  A_+^{'a} = A_+^a + \partial_-^{-1} \partial_\sigma A^{a\sigma}
\end{equation}
and integrate over this field combination. Simulating the effect
of the particles flying close to p$_B$ by an external source,
the modes of $A_+'$ with small $k_-$ could be used to describe
this source.

Performing the gaussian integral over $A_+^{'a}$ we obtain
\begin{eqnarray}
&S_1& = \int d^4x \lbrace {1 \over 2} A_{\sigma }^{a} \Box A^{a \sigma }
   \nonumber \\
&+& g f^{a b c } (A_{\sigma }^{a} \partial_{-} A^{b \sigma } )
{1 \over \partial_{-} }\partial_{\rho } A^{c \rho }
+ {g^2 \over 2} f^{a b c} f^{a b^{\prime } c^{\prime }   }
(A_{\sigma }^{b} \partial_{-} A^{c \sigma } ) {1 \over \partial_{-}^2 }
(A_{\rho }^{b^{\prime }} \partial_{-} A^{c' \rho })  \nonumber \\
&-& g f^{a b c } A_{\sigma }^{b} A_{\rho }^{c} \partial^{\sigma } A^{a
\rho }
-{g^2 \over 4 }f^{a b c } f^{a b^{\prime } c^{\prime }}
A_{\sigma }^{b} A_{\rho }^{c} A^{b^{\prime } \sigma } A^{c^{\prime }
\rho } \rbrace
\end{eqnarray}

The transverse components $A_\sigma$ are the physical gluon
fields. We use the complex notation for transverse vectors and
derivatives introduced above to write the action in the form
\begin{eqnarray}
  S_1 &=& \int d^4x \left\{ - \frac{1}{2} A^a \Box A^{a*} -
        i\frac{g}{2} \left[ A T^a
\stackrel{\leftrightarrow}{\partial}_- A^* \right]
       \cdot \partial^{-1}_- (\partial A^a
      + \partial^* A^{a*})  \right. \nonumber \\
 && \left. - \frac{g^2}{8} \left[ (AT^a
 \stackrel{\leftrightarrow}{\partial}_- A^*) \right] \partial^{-2}_-
 \left[ A T^a
    \stackrel{\leftrightarrow}{\partial}_- A^*  \right] \right.
      \nonumber \\
&&  + i\frac{g}{2}\left[AT^aA^*\right]\left(\partial A -
\partial^* A^* \left) -  \frac{g^2}{8} \left[ AT^a A^* \right]
\left[ AT^aA^* \right] \right\}   \right.
\end{eqnarray}
We divide the transverse gauge fields $A^a$ into parts with
respect to the corresponding momentum $k$
\begin{equation}
  A^a \rightarrow A^a + A^a_1 + {\cal A}^a  \;\;\;.
\end{equation}
${\cal A}$ represents the modes where
\begin{equation}
  k_+ k_- \ll k k^* \sim \mu^2_\bot \;\;\;.
\end{equation}
This is the typical momentum range for exchanged particles in
the peripheral scattering.

$A_1$ represents the modes where
\begin{equation}
  k_+ k_- \gg k k^* \sim \mu^2_\bot \;\;\;.
\end{equation}
These modes will be integrated out in the next step. We
integrate generally over all modes with $|\mbox{k}^2| \gg
\mu^2_\bot$. Formally our approximation does not include the
effects of a running coupling. They are not associated with
logarithms of the large energy $s$. However, from this it becomes
clear that in the resulting action the coupling is to be
understood as renormalized at a scale of order $\mu_\bot$, which
we assume to be large compared to the hadronic scale $\mu$.

The original notation $A$ is kept for the modes, where
\begin{equation}
  k_+ k_- \sim k k^* \sim \mu^2_\bot \; .
\end{equation}
This is the typical momentum range for scattering and produced
particles going in the $s$-channel direction. Moreover, the
dominante contribution comes from $s$-channel intermediate
states with particles strongly ordered and well separated in
longitudinal momenta according to the multi-Regge condition
(1.2) and close to mass shell, i.e.
\begin{equation}
  |k_+ k_- - k k^*| \ll \mu^2_\bot
\end{equation}
The effect of pairs of particles being not in the multi-Regge
configuration as well as the effect particles being off shell
(besides of the effects obtained by integrating over $A_1$
below) go beyond our approximation.
A systematic study of such corrections has been started in
\cite{FadLip}.

Now we analyze the terms resulting from the action (3.4) after
substituting the decomposition (3.5) with respect to momentum modes.

The kinetic term decomposes into three parts. By definition the
propagation between different modes is small. In the kinetic
term of ${\cal A}$ the transverse derivatives and in the one for
$A_1$ the longitudinal derivatives dominate
\begin{equation}
  \int d^4x \left\{  -2 A_1^a \partial_+ \partial_- A_1^{a*} - \frac{1}{2} A^a
    \Box A^{a*} + 2 {\cal A}^a \partial \partial^* {\cal A}^{a*}
    \right\}
\end{equation}
We concentrate now on the second term in (3.4). Its dominant
contribution comes from the configuration, where the field, on
which $\partial_-$ acts, has large $k_-$ and the field on which
$\partial_-^{-1}$ acts, has small $k_-$. From this we understand
that the form of this term with respect to $\partial_-$ leads
naturally to the ordering of $k_-$ and to the multi-Regge
kinematics (1.2).

Consider first the case if both fields entering the current $(A
T^a \stackrel{\leftrightarrow}{\partial}_- A^*)$ are not of type
${\cal A}$, i.e. decribe scattering particles. If also the third
field is not of type ${\cal A}$, then the resulting term
describes particle production by bremsstrahlung (Fig.~5a).
\begin{eqnarray}
-\frac{ig}{2}\left[ (A+A_1) T^a \partial_- (A + A_1)^* +
(A + A_1)^* T^a \partial_-
  (A + A_1) \right]   \nonumber \\
  \cdot\partial_-^{-1} \left[\partial (A^a +A^a_1)   +
    \partial^* (A^a + A^a_1)^*\right]
\end{eqnarray}
If the third field is of type ${\cal A}$ then the corresponding
term describes peripheral scattering of particles close to
p$_A$. We introduce \cite{KLS}
\begin{equation}
  {\cal A}_+ = - \frac{2}{\partial_-} \partial {\cal A} \;\;\;\;\;
  {\cal A}^*_+ = - \frac{2}{\partial_-}\partial^*{\cal A}^*
\end{equation}
and write the contribution as (Fig.~5b)
\begin{equation}
\frac{ig}{4} \left[ (A+A_1)  T^a \partial_- (A+A_1)^* + (A+ A_1)^* T^a
   \partial_- (A+A_1) \right] ({\cal A}^a_+ + {\cal A}_+^{a*})
\end{equation}
Consider now the case, where two fields are of type ${\cal A}$. Defining
 \cite{KLS}
\begin{equation}
  \partial^* {\cal A}^*_- = \partial_- {\cal A} \;\;\;\;
  \partial{\cal A}_- = \partial_-{\cal A}^*
\end{equation}
we have the contribution (Fig.~5c)
\begin{equation}
  \frac{ig}{2}((A_1 + A) T^a \partial {\cal A}_- + (A + A_1)^* T^a
     \partial^* {\cal A}^*_-) ({\cal A}^a_+ + {\cal A}_+^{a*})
\end{equation}
describing particle production. Typically ${\cal A}_-$ carries
a large longitudinal momentum component $k_-$, which is of the
same order as the one carried by $A$. The corresponding
momentum component carried by ${\cal A}_+$ is much smaller. The
multi-Regge chain with strongly ordered momentum components
$k_-$ appears by iterating this vertex. Due to the mass shell
condition (3.9) for fields of type~A the multi-Regge kinematics
(1.2) holds. The de\-fi\-nition of ${\cal A}_\pm$ takes into account
that the large momentum denominators have to cancel successively
against corresponding numerators. Using this notation we keep in
mind that the ${\cal A}_+$ and ${\cal A}_-^*$ are not independent.

There are more contributions from the second term in (3.4) where
the field entering with the derivative $\partial_-$ is of type
${\cal A}$.
\begin{eqnarray}
 && \left. ig\{\left[ \partial {\cal A}_- T^a (A+A_1) + \partial^* {\cal
A}_-^* T^a (A+A_1)^* \right] \partial_-^{-1} (\partial A^a +
\partial^* A^{a*})  \right. \nonumber \\
&& \left. + \frac{1}{2}\left[\left(\frac{\partial}{\partial_-}
{\cal A}_-\right)T^aA + \left(\frac{\partial^*}{\partial_-}{\cal
A}^*_-\right)T^aA^* \right] \left( \partial A^a + \partial^*A^{a*}
\right)\}  \right.
\end{eqnarray}
If the $k_-$ momentum component of the last field is much
smaller than those of the other fields, then the first term
contributes to the production of bremsstrahlung type, Fig. 5d,
and the second gives a small contribution. Here $( A + A_1 )$ is
to be replaced just by $A_1$, because the corresponding momentum
obeys (3.7).

In the other case, if the $k_-$ component of the last field in
(3.16) is of the same order as the one of ${\cal A}_-$, also the
second field in the bracket describes a quasi-real gluon (3.9) (
replace $( A + A_1 )$ by $A$ ). The result contributes to
scattering of particles close to $\mbox{p}_B$, with ${\cal A}_-$
describing the interaction with particles closer to $\mbox{p}_A$
( Fig. 5e). Then the second term is important and produces an
unpleasent  singularity in the multi-Regge limit, where for the
exchanged particles $k_-$ is much less than the transverse
momentum $k$. We return to the last case in a moment and show
that the singularity cancels.

Consider now the bremsstrahlung contribution of (3.16).
 The large longitudinal
momentum in the denominator of the $A_1$ propagator cancels, if
the adjacent vertex radiates bremsstrahlung. The corresponding
contributions to particle production are of the same order as
the one by the vertex (3.15) and adding them the latter vertex
turns into the effective vertex. Clearly only one $A_1$
propagator is compensated by one bremsstrahlung vertex.
Contributions with more $A_1$ propagators are suppressed by
powers of large sub-energies in the multi-Regge kinematics.
Therefore we integrate over $A_1$ just by picking up the
bremsstrahlung contributions and including them into the nearest
effective production vertex. In this way we are led to consider
the contributions to particle production shown in Fig.~6.

The sum of the first two terms (in both cases) has the same
colour structure as the third term. In momentum space the third
graph is in both cases proportional to
\[
  g \frac{2}{|q|^2} \cdot \left( \frac{q^*}{2} A + \frac{q}{2} A^*
  \right) \nonumber
\]
The sum of the first two bremstrahlung terms yields correspondingly the
momentum factors
\begin{equation}
  -g \frac{p_-}{(\mbox{p}+\mbox{k})^2} \frac{1}{k_-} \left( k^* A +
   k A^* \right) = -g \left( \frac{1}{k} A + \frac{1}{k^*}
   A^*\right)
\end{equation}
which can be related to the effective bramstrahlung vertex
\begin{equation}
-\frac{ig}{2}(\partial\partial^*{\cal A}^a_-)
(\frac{1}{\partial^*}A ) T^a ({\cal A}_+ + {\cal A}^*_+ )
 \;\;+\;\; \mbox{c.c.} \;\;.
\end{equation}
We have used (3.7) for the $A_1$ propagator, allowing the
approximation \linebreak (p+k)$^2 \approx p_- k_+$, and the mass-shell
condition (3.9) for the momentum $k$ of the produced gluon. The
resulting contribution corresponds to replacing the particle
production vertex (3.14) by the following effective vertex being
the sum of expressions (3.14) and (3.18)
\begin{equation}
  - \frac{ig}{2}\{\left[ (\partial^{*-1} A) T^a (\partial {\cal A}_-) \right]
    \partial^* ({\cal A}^a_+ + {\cal A}^{a*}_+) +
    \left[(\partial^{-1} A^*) T^a (\partial^* {\cal A}^*_-)\right] \partial
    ({\cal A}^a_+ + {\cal A}^{a*}_+)\}
\end{equation}
There is no contribution from bremsstrahlung to the production
of particles belonging with respect to longitudinal momentum
ordering (1.2) not to the adjacent vertex. This would take more
than one $A_1$ propagator and is suppressed as discussed above.

This is also the reason why the third term in (3.4) does not
contribute to particle production. This term gives no essential
contribution with one or more of the involved fields being of
the type $A_1$. Therefore it was possible to do the $A_1$
integral without mentioning about this term.

When in the third term of Eq.(3.4) two fields are of the type
${\cal A}$, this results in terms of the form
\begin{eqnarray}
  &\mbox{}& \left[ A T^a (\partial {\cal A}_-) + A^* T^a (\partial^* {\cal
A}^*_-) \right] \partial_-^{-2} \left[ A T^a (\partial_-^2
\partial^{*-1} {\cal A}^*_+) + A^* T^a (\partial^2_- \partial^{-1}
{\cal A}_+) \right] \nonumber \\
  &\mbox{}& \mbox{or}  \\
  &\mbox{}&  \left[ A T^a (\partial {\cal A}_-) + A^* T^a (\partial^*
{\cal A}^*_-) \right] \partial_-^{-2} \left[ (\partial_- A) T^a
(\partial_- \partial^{*-1} {\cal A}^*_+) + (\partial_- A^*) T^a
(\partial_- \partial^{-1} {\cal A}_+) \right]  \nonumber
\end{eqnarray}
In order to compensate the denominator in the first case the
particle corresponding to ${\cal A}_+$ has to have a large $k_-$
component compared to the one of $A$ in the second bracket. Then
the propagator of ${\cal A}_+$ would have a large longitudinal
part, i.e. the modes are in fact of type $A_1$ rather than of
type ${\cal A}$. Therefore the first term in (3.20) gives a
negligible contribution. In the second case the large
denominator should be cancelled by the momenta of both fields in
the second bracket. But only one of them can have a large $k_-$ component.

We see that the only non-negligible contribution of the third
term in (3.4) arises in the case if all fields are of type A
(corresponding to elastic scattering) or in the case if one
field is ${\cal A}_-$ and the others are of type A
(corresponding to the ``right end'' of a multi-Regge chain)
(Fig.~7)). Both contributions are represented by the vertex
\begin{equation}
\frac{ig}{2} \left( {\cal A}^a_- + {\cal A}^{a*}_-
\right)\partial\partial^* \partial^{-2}_-\left[AT^a
\stackrel{\leftrightarrow}{\partial}_- A^* \right]
\end{equation}
Indeed, the integral over ${\cal A}_{\pm}$ with the kinetic term
(3.10) and the vertices (3.13) or (3.15) (omitting $A_1$ in
those expressions ) and (3.21) reproduces the mentioned
contributions of the third term in (3.4). The vertex (3.21) can
be understood as describing the scattering of produced particles
close in momenta to the right incoming particle with momentum
$\mbox{p}_B$. There are more contributions of this type, Fig.~5e,
arising in particular from the second term in (3.4), if one of
the fields is ${\cal A}_-$. Such a contribution is contained in
(3.16) as discussed above, if the second field $(A + A_1)$ in
the bracket is of type $A$. Another contribution is contained in
(3.13) : we express the field of type ${\cal A}$ as ${\cal A}_-$
(3.14) instead of ${\cal A}_+$ and arrive at
\begin{equation}
-\frac{ig}{2}[\partial^{-2}_-\partial\partial^*\left({\cal A}^a_- + {\cal
A}^{a*}_- \right)]\left[AT^a\stackrel{\leftrightarrow}{\partial}_-A^*\right]
\end{equation}
This contribution cancels the one of (3.21) from the first term
in (3.4). Therefore we are left with the contribution from
(3.16). Omitting $A_1$ we rewrite it up to the total derivatives
as
\begin{eqnarray}
&& \left. -ig\{{\cal A}^a_-\partial \left[AT^a\partial^{-1}_-\left(
\partial A + \partial^* A^* \right)\right]  \right. \nonumber \\
&& \left. + \frac{1}{2}\left(\partial^{-1}_- {\cal A}^a_-
\right) \partial \left[ AT^a\left( \partial A + \partial^* A^*
\right)\right]\}  + \;\;\; \mbox{c.c.}  \right.
\end{eqnarray}
Integrating over $A_1$ we took into account up to now the
particular contribution ( Fig.~6), if in one of the adjacent
vertices there is a type ${\cal A}$ field. There are
corresponding contributions from the case, if this type ${\cal
A}$ field is replaced by a type $A$ one and also if both
adjacent vertices have a type ${\cal A}$ field. They result in
vertices, which are readily obtained from the bremsstrahlung
part (3.18) of the effective vertex (3.19):  \\

$\;\;\;$ in the first case by substituting ${\cal A}_+$ by $A$
via (3.12)
\begin{equation}
-ig\frac{1}{2}\left( \partial^{* -1}AT^a\partial\partial^*{\cal
A}_- \right) \cdot \partial^{-1}_- \left( \partial A^a +
\partial^* A^{a*} \right) + \;\; \mbox{c.c.}
\end{equation}
$\;\;\;$ and in the second case by replacing $A$ by ${\cal A}_+$
\begin{equation}
-\frac{ig}{4}\frac{1}{2}[\partial_-
{\left(\partial\partial^*\right)}^{-1}{\cal
A}_+]T^a\left(\partial\partial^*{\cal A}_-\right)\left({\cal A}^a_+ +
{\cal A}^{a*}_+ \right) + \;\;\;\mbox{c.c.}
\end{equation}
There is an additional factor $\frac{1}{2}$ since now two fields
are of the same type. Indeed, the first two graphs in Fig.~6 are
equivalent, if the full and dashed lines are not distinguished.

The fourth term in (3.4) does not involve longitudinal
derivatives and may seem to be unimportant in the considered
asymptotics. Neverteless it gives rise to singular contributions
compensating the ones in (3.23) (compare \cite{KLS}).
Indeed, its contribution in the
case if one of the fields is ${\cal A}_-$ is given by
\begin{equation}
 \frac{ig}{2}\left[( \partial^*\partial^{-1}_-{\cal A}^*_-) T^aA^* +
AT^a(\partial\partial^{-1}_-{\cal A}_-)\right] \left( \partial A^a
- \partial^*A^{a*} \right)
\end{equation}
and can be represented up to total derivatives as
\begin{equation}
\frac{ig}{2}\left( \partial^{-1}_-{\cal
A}^a_- \right) \partial \left[ AT^a \left( \partial A - \partial^*
A^* \right) \right] + \;\;\; \mbox{c.c.}
\end{equation}

We sum the contributions (3.23),(3.24) and (3.27) and obtain
using (3.9) the scattering vertex for particles close to $\mbox{p}_B$
\begin{equation}
\frac{ig}{2} A^a_- \left[(\frac{\partial}{\partial^*}A) T^a
(\partial_+ \frac{\partial^*}{\partial}A^*) \right] + \;\;\; \mbox{c.c.}
\;\;\;.
\end{equation}
In Eq.(3.28) we denoted by $A_-$ the field ${\cal A}_-$
restricted to the real values.
We observe, that in the dominating interaction terms, ${\cal A}_+$
and ${\cal A}^*_+$ enter always as the sum. Therefore it is
consistent to relax the constraint relating ${\cal A}_+$ and
${\cal A}^*_-$ and to consider ${\cal A}_+$ and ${\cal A}_-$ as
real and independent ( denoted by $A_+$ and $A_-$, respectively).
 This has also been used to obtain (3.28).

In the gauge $A_- = 0$ with the momentum $\mbox{p}_B$ of the
right incoming particle as the gauge vector ( R - gauge ) the
current of the scattering particles close to $\mbox{p}_A$ has a
simple form (3.13),
\begin{equation}
   \tilde{j}^a_- = i\left( AT^a\stackrel{\leftrightarrow}{\partial}_-
A^*\right) \;\;,
\end{equation}
whereas the current of scattering particles close to
$\mbox{p}_B$ is obtained from a sum of many contributions as (3.28)
\begin{equation}
\tilde{j}^a_+ = i \left( A_L T^a
\stackrel{\leftrightarrow}{\partial}_+ A^*_L \right) \;\;\; .
\end{equation}
$A_L$ represents the transverse components of the gauge
potential in the L-gauge, $A_+ = 0$, related to $A = A_R$ by
\begin{equation}
\partial A_L = - \partial^* A^*
\end{equation}

The result (3.30) was to be expected, because of the parity
symmetry interchanging incoming particles A and B . The
effective vertex (3.19) is symmetric under this transformation
because of (3.31). Parity symmetry of the scattering terms
determines the form (3.30) of $\tilde{j}^a_+$ if
$\tilde{j}^a_-$ (3.29) is given.

Let us discuss the argument to derive (3.31) and return to the
gauge field action (3.1) before the integration over $A_+'$. Any
correlation function with $\partial_\mu A^\mu$ inserted is small
${\cal O} (g)$
 due to the form of integral over $A'_+$. So in both the R
rauge $(A_- =0)$ and the L gauge $(A_+ = 0)$ we have
correspondingly $(A_R = A)$
\begin{equation}
  \partial_\mu A^{a\mu}_R = {\cal O}(g) \quad , \quad
  \partial_\mu A_L^{a\mu} = {\cal O}(g)
\end{equation}
{}From the gauge transformation relating $A_R$ and $A_L$ we have
the following relation between the transverse components written
in complex notation
\begin{equation}
  \partial A^a_L = \partial A^a_R + \partial \partial^* \omega^a
+ {\cal O}(g)
\end{equation}
Here $\omega^a = -\partial_-^{-1} A^a_{L-} = \partial_+^{-1}
A^a_{R+}$ is the parameter of the gauge transformation. Together
with (3.32) this leads to (3.31) up to ${\cal O}(g)$. This
relation corresponds to the one between the polarization vectors
(2.4). It suggests to introduce a complex scalar field $\phi^a$
replacing the transverse components of the gauge potential in
describing the scattering gluons \cite{Leff}:
\begin{equation}
  A^a = A_R^a = i \partial^* \phi^a \; , \quad A_L^{a*} = i
\partial^* \phi^{a*} \;\;\;.
\end{equation}

We have shown that the dominating contributions of the
interaction terms in (3.4) can be represented by the effective
gluon production vertex (3.18) and by vertices involving the
currents $j_{-}$ (3.30) and $j_{+}$ (3.29) of scattering gluons close
to p$_A$ and p$_B$, respectively.
\begin{eqnarray}
  S^{eff}_g &=& S_{ksg} + S_{kpg} + S_{sg} + S_{pg} \nonumber \\
  S_{ksg} &=& - \frac{1}{2} \int d^4x A^a \Box A^{a*}, \;
        S_{kpg} = -2 \int d^4x  A^a_+ \partial \partial^*
       A^a_- \nonumber \\
  S_{sg} &=&  \frac{g}{2} \int d^4x \left\{ \tilde{j}^a_{-} A^a_+
       + \tilde{j}^a_{+}  A^a_- \right\} \nonumber \\
  S_{pg} &=& ig \int d^4x \left\{ J^{a*} (\partial^{*-1} A^a) -
         J^a (\partial^{-1} A^{a*}) \right\} \nonumber \\
   J^{a*} &=& \partial^* A_+ T^a \partial A_-
\end{eqnarray}
Note, that because we consider the fields $A_\pm$ as
being independent ones the coefficient in $S_{kpg}$ differs from
one which we get by substitution of definitions (3.12) and
(3.14) into (3.10) \cite{KLS}.
The appearance of $A_L$ in $\tilde{j}_+$ is essential, because this guarantees
symmetry under interchanging the incoming particles p$_A$
and p$_B$. The symmetry of the production term holds because of (3.31).
Starting the analysis with particles close to p$_B$ and working
in L-gauge $A_+ = 0$ we obtain the same result.

We stress that the fields appearing in the result (3.35) are
just different modes of one and the same gluon field A. It is
obtained just by rewriting (3.4) and integrating over modes with
large longitudinal momentum components (3.7) with approximations
keeping the dominant contributions to high-energy peripheral scattering.

We did not include the vertex (3.25) involving three particles
of the type ${\cal A}$ and its counterpart obtained by
interchanging indices $+$ and $-$ . These vertices do not contribute
to tree amplitudes, and therefore they did not arise in the
graphical approach in chapter~2.
But they give rise to reggeization of exchanged particles as
well as $s$-channel intermediate states with virtual scattering
particles. The sum of both contributions to reggeization is
independent  of the parameter $\mu_{\perp}$ introduced to
separate $A_1$ and A. If we do not restrict the virualness of
the scattered particles A between vertices involving ${\cal
A}_{\pm}$ then the latter contribution yields the complete
reggeization.  Under this assumption vertices of the type (3.25)
do not appear in the effective action. We plan to discuss this
point involving the bootstrap relations in a separate publication.

By introducing $\phi^a$ (3.34) we resolve the constraint
relating $A = A_R$ and $A_L$ and obtain (3.29) with the substitutions
\begin{eqnarray}
  S_{ksg} &=& - \frac{1}{2} \int d^4x \partial \phi^a \Box
      \partial^* \phi^{a*} \nonumber \\
  S_{pg} &=& - g \int d^4x \left\{ J^{a*}
  \phi^a + J^a \phi^{a*} \right\}    \nonumber \\
  \tilde{j}^a_{-} &=& i [ \partial^* \phi T^a \partial_- \partial \phi^* +
      \partial \phi^* T^a \partial_- \partial^* \phi]  \nonumber \\
  \tilde{j}^a_{+} &=& i[ \partial^* \phi^* T^a \partial_+ \partial \phi +
      \partial \phi T^a \partial_+ \partial^* \phi^*]
\end{eqnarray}
Now the analysis can be generalized to the case with quarks
included. We decompose the quark field $\psi^a$ into light-cone
components (see Appendix)
\begin{eqnarray}
\psi^a = \psi_{-}^a + \psi_{+}^a,\ \ \psi_{-}^a ={1 \over 4} \gamma_{+}
\gamma_{-} \psi^a , \ \ \psi_{+}^a = {1 \over 4 }
\gamma_{-} \gamma_{+} \psi^a  .
\end{eqnarray}
We choose the gauge $A_{-}^a = 0 $ and obtain for the fermionic terms of
the QCD action
\begin{eqnarray}
&i& \overline \psi \gamma^{\mu }(\partial_{\mu } - i g  t^{a} \mbox{A}_{\mu
}^{a}) \psi =
i \overline \psi_{+}^{\prime } \gamma_{+} \partial_{-} \psi_{+
  }^{\prime }
+{i \over 4} \overline \psi_{-}  \gamma_{-} {\Box \over
\partial_{-} } \psi_{-} \cr
&+&{g\over 2} (\overline \psi_{-} t^{a} \gamma_{-} \psi_{-} ) A_{+}^{a}
+ g (\overline \psi_{+}^{\prime } t^{a} \gamma_{\sigma } \psi_{-} +
\overline \psi_{-} t^{a} \gamma_{\sigma } \psi_{+}^{\prime } ) A^{a
\sigma }   \cr
&-&{g\over 4}A^{a\sigma}[
(\frac{\partial^{\rho}}{\partial_-}  \overline \psi_{-})
 \gamma_{\rho} \gamma_{-}
 \gamma_{\sigma }t^a \psi_-  +
 \overline \psi_- \gamma_{\sigma } \partial_{-}^{-1} \gamma_{-}
\hat \partial_{\perp }  t^a \psi_{-}]
\end{eqnarray}
We introduced
\begin{equation}
\psi_{+}^{\prime a} = \psi_{+}^a + \partial_{-}^{-1} { \gamma_{-} \over 4 }
\hat \partial_{\perp }  \psi_{-}^a
\end{equation}
and used the notations
\begin{eqnarray}
\Box = 4 \partial_{+} \partial_{-}  + \Box_{\perp },
\Box_{\perp } = \partial_{\sigma } \partial^{\sigma },
\hat \partial_{\perp } = \gamma_{\sigma } \partial^{\sigma }.
\end{eqnarray}
We assumed also the summation over the fermion colour indices which are
suppressed for simplicity of notation.

Now we see that $\psi'_+$ plays a role analogous to $A'_+$
above. Performing the gaussian integration over $\psi'_+$ and
$A'_+$ we are left with $A$ and $\psi_-$. Analogous to (3.5) we decompose
$\psi_-$ into modes corresponding to the kinematic regions (3.7),
(3.9) and (3.6), respectively.
\begin{equation}
  \psi_- \rightarrow \psi_{1-} + \psi_- + \tilde{\psi}_-
\end{equation}
Integration over $A_1$ and $\psi_{1-}$ leads to effective
vertices of gluon and fermion production. Analogous to ${\cal
A}_+$ we introduce
\begin{equation}
  \tilde{\psi}_+ = - \partial_-^{-1} \frac{\gamma_-}{4} \hat{\partial}_\bot
        \tilde{\psi}_- \;\;\;\; \bar{\tilde{\psi}}_+ = -\frac{1}{4}
                \left(\frac{\partial^{\rho}}{\partial_-}
                \bar{\tilde{\psi}}_-\right)\gamma_{\rho}\gamma_-
\end{equation}
and the analogon of ${\cal A}_-$ is $\tilde{\psi}_-$ itself.

After the integration over $\psi'_+, A'_+$ with (3.35) the
fermionic kinetic term becomes
\begin{equation}
  \int d^4x \left\{ \frac{i}{4} \bar{\psi}_{1-} \gamma_-
\partial_+ \psi_{1-} + \frac{i}{4} \bar{\psi}_-
\gamma_- \frac{\Box}{\partial_-} \psi_- + i \bar{\tilde{\psi}}_+
  \hat{\partial}_\perp \tilde{\psi}_-  + i \bar{\tilde{\psi}}_-
  \hat{\partial}_\perp \tilde{\psi}_+  \right\}
\end{equation}
Despite of the common origin (see Eq.(3.42)), the fields
$\tilde{\psi}_+ , \tilde{\psi}_-$ in Eq.(3.43) are treated as
being independent ones. This again leads to the coefficients in front
of terms involving these fields being different than the
coefficients which follow from the substitution of
Eqs.(3.41),(3.42)  into (3.38) \cite{KLS}.

It is convenient to decompose the fermion fields with respect to
a basis of Majorana spinors $u_{ij} \; (i, j = +,-)$ (see Appendix).
\begin{eqnarray}
\psi_{-} &=& 2i[(\partial \chi_{-})u_{- +} -
   (\partial^* \chi_{-}^{*})u_{- -}]\;\; , \ \ \
\psi_{+} = 2i[(\partial\chi_{+}) u_{+ +} -
 (\partial^* \chi_{+}^{*})u_{+ -}]\;\; ,  \cr
\tilde{\psi}_{-}  &=& a_{-}^{*} u_{- +} +
a_{-} u_{- -}\;\;, \; \; \; \;
\tilde{\psi}_{+}  = a_{+}^{*} u_{+ +} +
a_{+} u_{+ -}\;\;, \nonumber \\
\bar{\psi}_- &=& -2i[(\partial^*\bar{\chi}_-)\bar{u}_{- +} -
(\partial\bar{\chi}^{*}_-)\bar{u}_{- -}]\;\;, \ \ \
\bar{\psi}_+ = -2i[(\partial^*\bar{\chi}_+)\bar{u}_{+ +} -
(\partial\bar{\chi}^{*}_+)\bar{u}_{+ -}]\;\;, \nonumber \\
\bar{\tilde{\psi}}_{-}  &=& \bar{a}_{-}^{*} \bar{u}_{- +} +
\bar{a}_{-} \bar{u}_{- -}\;\;, \; \; \; \;
\bar{\tilde{\psi}}_{+}  = \bar{a}_{+}^{*} \bar{u}_{+ +} +
\bar{a}_{+} \bar{u}_{+ -}
\end{eqnarray}
The scattering quarks are now described by the complex-valued
fields ${\chi}_-$ and ${\chi}^*_-$. In the case of
Dirac fermions they are not related to each other by complex
conjugation, but such a relation holds in the case of Majorana
fermions.

Starting the analysis
with particles close to p$_B$ and working in the L gauge $A_+ =
0$ we would describe the scattering fermions instead by $\psi_+$
with components ${\chi}_+$ and ${\chi}^*_+$. Their
relation to ${\chi}_-$ and ${\chi}_-^*$ is obtained
in an analogous way as we obtained the relation (3.31) between
$A_R = A$ and $A_L$. Temporarily we write $\psi_R$ for the field
in the R gauge, $A_- = 0$, and $\psi_L$ for the field in the L
gauge, $A_+ =0$. The form of the integral over $\psi'_+$ in the
R gauge or the analogon $\psi'_-$ in the L gauge leads to the  relation
\begin{equation}
  \partial_- \psi_{R+} + \frac{1}{4} \gamma_-
\hat{\partial}_\bot \psi_{R-} = {\cal O}(g), \partial_+ \psi_{L-} + \frac{1}{4}
  \gamma_+ \hat{\partial}_\bot \psi_{L+} = {\cal O}(g)
\end{equation}
The gauge transformation relating the two descriptions yields
$\psi_L = \psi_R + {\cal O}(g)$ and this implies
\begin{equation}
  \partial_- {\chi}_+ = -\partial^* {\chi}^*_-\;\;, \;\;\;
  \partial_+{\chi}_- = -\partial^* {\chi}^*_+ \;\;.
\end{equation}
The complex conjugation of Eq.(3.46) leads to the relations
between fields with bar ($\bar{\chi}$, $\bar{\chi}^*$).

The modes of type ${\cal A}$ in the components $a_\pm$,
$a^*_\pm$ ( also $\bar{a}_\pm$, $\bar{a}^*_\pm$)
describe the exchanged fermions. Originally they are
not independent since  there is a relation
between fields with index plus and those with index minus from
(3.42). As we already mentioned above, this constraint is
relaxed \cite{KLS} and we consider
$a_\pm$, $a^*_\pm$ and those fields with bar as being
independent ones. Only for Majorana particle we have $\bar{a} =
a^*$ and $\bar{a}^* = a$.
 Similar to the case of ${\cal A}_\pm$
this turns out to be consistent and correctly accounts for the
independent degrees of freedom.

Instead of repeating in detail the analysis of the interaction
terms in the case of fermions included we show in the next
section, how supersymmetry can be used to reconstruct the
fermionic terms from the known gluonic ones. For this we
temporarily restrict ourselves to one Majorana field and change
its gauge group representation to the adjoint
one $(t^a \rightarrow T^a)$.

\section{ Reconstruction by supersymmetry relations }
\setcounter{equation}{0}
\hspace*{4mm} We use the supersymmetry in Yang-Mills theory with a Majorana
fermion in the adjoint representation.
The action has a form
\begin{equation}
  S_{YM} = \int d^4 x \left\{ - \frac{1}{4} \mbox{F}^a_{\mu\nu}
         \mbox{F}^{a_{\mu\nu}} + \frac{i}{2} \bar{\psi}^a
         (\hat{\partial}_{ab} - ig T^c_{ab}
         \mbox{\^{A}}^c) \psi^b \right\}
\end{equation}
where the spinor $\psi^a$ obeys the Majorana condition,
  \[ \psi^a = C \bar{\psi}^{aT} \quad , \qquad \mbox{and} \qquad
    (T^a)_{bc} = (-i) f^{abc} \; . \]
(4.1) is invariant under the transformation
\begin{eqnarray}
  \delta\mbox{A}^a_\mu &=& 2i \bar{\alpha} \gamma_\mu \psi^a = - 2i
         \bar{\psi}^a \gamma_\mu \alpha \nonumber \\
  \delta \psi^a &=& \sigma_{\mu\nu} \mbox{F}^{a\mu\nu} \alpha \\
  \sigma_{\mu\nu} &=& \frac{1}{2} [ \gamma_\mu ,\gamma_\nu ] \nonumber
\end{eqnarray}
but the algebra of transformations (4.2) does not
close on translations without introducing auxiliary fields \cite{West}.

We restrict ourselves to the subset of transformations (4.2) by choosing
the parameter $\alpha$ in the form (see Appendix)
\begin{equation}
  \alpha = \alpha_+ u_{+-} + \alpha^*_+ u_{++}
\end{equation}
with $\alpha_+$ being a grassmanian complex number.
In this case the algebra of transformation (4.2) restricted by condition
(4.3) closes on the translations $(\partial_-)$ as is required by
supersymmetry \cite{supers}. Moreover this subset of transformations
leaves $A^a_-$ invariant and therefore it is appropriate for
working in the R-gauge $A^a_- = 0$.

Under this subset of transformations (4.2) the transverse gauge
fields $A^a_\sigma$ and the fermion field projection $\psi_-$
transform into each other i.e.
\begin{eqnarray}
  \delta A^a_\sigma &=& 2i \bar{\alpha} \gamma_\sigma \psi^a_-
        \nonumber \\
   \delta \psi_-^a &=& - 2 \partial_- (\gamma^\sigma
           A^a_\sigma)\gamma_+ \alpha
\end{eqnarray}
Above, in chapter 3, we have introduced the fields $\phi^a$ and $\chi^a_-$
(Eqs. (3.34), (3.44)) replacing the transverse
components $A^a_\sigma$ of the gauge fields and the spinor field
$\psi^a_-$. The transformation law of fields $\phi^a$ and
$\chi^a_-$ read
\begin{eqnarray}
   & \delta \phi^a = 4 i\alpha_+ \chi^{a*}_-  &  \delta \phi^{a*}
        =  4i \alpha^*_+ \chi^a_- \nonumber \\
   & \delta \chi^a_- = -2 (\partial_- \phi^{a*}) \alpha_+ & \delta
          \chi_-^{a*} = - 2 (\partial_- \phi^a) \alpha_+^*
\end{eqnarray}
Disregarding contributions ${\cal O}(g)$ we find that also fields
$A'^a_+$ and $\psi'^A_+$ (Eqs. (3.2) and (3.39)) transform into
each other
\begin{eqnarray}
  \delta A'_+ &=& 2i \bar{\alpha} \gamma_+ \psi'_+ \nonumber \\
  \delta \psi'_+ &=& -2 (\partial_- A'_+) \alpha
\end{eqnarray}
These fields are integrated out and we have separated the modes
of the remaining fields $A_\sigma$ and $\psi_-$ with respect to
kinematics. For the modes of type A, related to the scattering
particles close to p$_A$, we have the supersymmetry
transformation given above (4.5). For the modes of type ${\cal
A}$ the transformations follow also from (4.4) by their
definition in terms of $A_\sigma, \psi$.
\begin{eqnarray}
   A_{\pm} &=& \frac{1}{2}[{\cal A}_{\pm} + {\cal A}^*_{\pm}]
\nonumber \\
  \delta A_+ &=& 2i[\alpha^*_+ a_+ + \alpha_+ a^*_+] \nonumber \\
  \delta A_- &=& -i\left[\alpha^*_+
\left(\frac{\partial_-}{\partial} a^*_-\right) + \alpha_+\left(
\frac{\partial_-}{\partial^*}a_-\right)\right] \nonumber \\
   \delta a_+^* &=& - 2 (\partial_- A_+) \alpha^*_+
\nonumber \\
   \delta a^*_- &=&  4 \alpha_+ (\partial A_-)
 \end{eqnarray}
 We repeat that in Eq.(4.7), the fields $A_{\pm}$ are defined by the
transverse part of gluon potential and should not be confused
with the corresponding fields in original QCD lagrangian (3.1).
 Similar as in (4.5) the remaining transformation follow by
complex conjugation.

The kinetic term of the modes of type A has the form
\begin{eqnarray}
  S_{ks} &=& \int d^4 x \left[\frac{1}{2} A^a_\sigma \Box
      A^{a\sigma} + \frac{i}{8} \bar{\psi}^a_- \gamma_-\frac{1}{\partial_-}
       \Box \psi^a
      \right] = \nonumber \\
      &=& \int d^4 x \left[ - \frac{1}{2} (\partial^* \phi^a) \Box
      (\partial \phi^{a*}) + i (\partial \chi^a_-) \frac{1}{\partial_-}
      \Box (\partial^* \chi^{a*}_-) \right]
\end{eqnarray}
It is invariant under transformations (4.5).

The minimal coupling $\partial_+\delta_{bc} \rightarrow \partial_+\delta_{bc} -
\frac{ig}{2}A^a_+T^a_{bc}$
in Eq. (4.8) leads to the interaction
term (compare (3.36))
\begin{equation}
 \frac{g}{2} \int d^4 x \left(  j_-^a + j^{a*}_- \right) A^a_+
\end{equation}
with the current $j^a_-$ given by
\begin{equation}
  j^a_- = -i (\partial_- \partial^* \phi) T^a (\partial \phi^*) +
         2 (\partial \chi_-) T^a (\partial^* \chi^*_-)
\end{equation}
Now, applying the transformations (4.5) to the current $j^a_-
(j_-^{a*})$ we obtain its superpartner $v_-^{a*}(v_-^a)$
\begin{eqnarray}
  \delta j_-^a = 2 \alpha_+ \partial_- v_-^{a*} \qquad \delta v_-^{a*}
       = - 4 i \alpha^*_+ j^a_- \nonumber \\
       v_-^{a*} = - 2 (\partial \phi^*) T^a (\partial^* \chi ^*_-)
\end{eqnarray}
The interaction terms of the scattered particles are obtained by
constructing the invariant $S_{s-}$ with respect to the
transformations (4.5) and (4.7) and involving the currents
$j^a_-, v^a_-$ and fields $ A^a_+, a^a_+$. It is given
by the formula
\begin{equation}
  S_{s-} =  \frac{g}{2} \int d^4 x \left[ (j^a_- + j^{a*}_-) A^a_+
       + v_-^{a*} a^a_+  -  v^a_-a^{a*}_+ \right]
\end{equation}
Under the transformations (4.5),(4.7) the integrand of (4.12)
changes by the total $\partial_-$ derivative ( we use here that
in our kinematics $\partial_- a_+$ is small). \\
In terms of component fields the action (4.12) takes the form
\begin{eqnarray}
  S_{s-} &=& -  g \int d^4 x \left[ \frac{i}{2}
       (\partial_- \partial^* \phi) T^a (\partial \phi^*) A^{a*}_+
       \right.  \nonumber \\
   && \left. - (\partial \chi_-) T^a (\partial^* \chi^*_-)
      A^{a*}_+ - (\partial^* \chi^*_-) T^a (\partial \phi^*)
      a^a_+  + \mbox{h.c.} \right]
\end{eqnarray}

Supersymmetry connects also the terms in the effective action
describing production of gluons and fermions.
We have seen that only one combination of the gluon fields
$(\partial {\cal A} + \partial^*{\cal A}^*) \sim A_\pm$
transfers the interaction between scattering particles. However
there are fermion exchanges of two types, $a_\pm$ and $a^*_\pm$,
distinguished by the flow of helicity. Therefore the
reconstruction of fermionic terms related to particle exchanges
and production is not straightforward if we start from the
leading gluonic terms. The considered supersymmetry subgroup
produces leading fermionic terms both from leading and
non-leading gluonic ones. There is a simple supersymmetry
relation between the exchanged gluons $A_\pm$ and the exchanged
fermions with one helicity only. It is different from relations
(4.7) derived for $A_\pm , a_\pm , a^*_\pm$ from their
definitions.

The corresponding transformations leave the kinetic term of the
exchanged fields invariant if only one of the two fermionic
terms is kept (see Eqs.(3.43) and (3.44))
\begin{equation}
\tilde{S}_{kp} = \int d^4x \left\{ - 2A^a_+\partial\partial^* A^a_-
- ia^{a*}_+\partial^* a^{a*}_- \right\}
\end{equation}
The transformations are
\begin{eqnarray}
\tilde{\delta}A_+ = 2i\alpha_+a^*_+ \;\; &,& \;\;\;
\tilde{\delta}A_- = -2i\alpha_+\frac{\partial_-}{\partial}a^*_-
\;\;, \nonumber \\
\tilde{\delta}a^*_+ = -4\alpha^*_+(\partial_-A_+) \;\;\; &,&
\;\;\; \tilde{\delta}a^*_- = 4\alpha_+(\partial A_-)
\end{eqnarray}
We write the gluonic production vertex as $-g(J^a\phi^{a*} +
J^{a*}\phi^a)$ and we extend the form of the effective current
$J^a$ in such a way, that its variation under (4.15) is a total
$\partial_-$ derivative
\begin{equation}
J^{a*} = -[(\partial A_-)T^a(\partial^*A_+) - \frac{i}{2}(\partial^*
a^*_+)T^aa^*_-]\;\;.
\end{equation}
We obtain the multiplet of effective currents
\begin{eqnarray}
  \tilde{\delta} J^{a*} &=& 2\alpha^*_+(\partial_-v^a) \nonumber  \\
    \tilde{\delta} v^a &=& -4i\alpha_+ J^{a*}
\end{eqnarray}
where
\begin{equation}
v^a = -ia^*_-T^a(\partial^*A_+)  \;\;\;.
\end{equation}
{}From Eqs.(4.5) and (4.17) we see that
\begin{equation}
  -g\int d^4x (\phi^aJ^{a*} - \chi^{a*}_-v^a)
  \end{equation}
 is invariant.

    We write down
this invariant explicitly and we add to it analogous terms
corresponding to the other fermion helicity
\begin{eqnarray}
  S_{pg} + S_{p-} &=& - g \int d^4 x \left[ \phi^{a*}\left( (\partial^*
A_-) T^a (\partial A_+) - \frac{i}{2} a_- T^a (\partial a_+ )
\right) \right.  \nonumber \\
&+& \left. i\chi^a_- a_- T^a (\partial A_+) + \mbox{h.c.} \right]
\end{eqnarray}
Analogously the kinetic term for exchanged particles is obtained
from (4.14) by adding the contribution of the other fermion
helicity. It follows directly from (3.43) and (3.44)
\begin{equation}
S_{kp} = \int d^4x \left\{ -2A^a_+\partial\partial^* A^a_-
-ia^{a*}_+\partial^*a^{a*}_- -ia^a_+ \partial a^a_- \right\} \;\;\;.
\end{equation}

As we have seen the supersymmetry (4.3), (4.4) allows easily to
reconstruct the scattering and production terms involving
fermions if we restrict ourselves to particles close to p$_A$.
Fermion exchange over a large rapidity interval is suppressed
and of the same order as small contributions neglected in course
of the derivation above. The easiest way to reconstruct the
missing fermionic terms for particles close to p$_B$ is to
repeat the procedure from the beginning, i.e. concentrating on
particles close to p$_B$ and working in R gauge $A_+ = 0$. The
results are analogous with the substitutions of indices $+
\leftrightarrow -$ and $\phi \leftrightarrow \phi^*$. To obtain
the full effective action with one Majorana fermion we have to
add to the gluonic action (3.36) the fermionic kinetic terms
from (4.8) and (4.21), the fermionic interaction terms from
(4.13) and (4.20) and further fermionic interaction terms
$S_{s+}$ and $S_{p+}$ obtained from the latter $S_{s-}$ and
$S_{p-}$ by the substitutions $+ \leftrightarrow -, \; \phi
\leftrightarrow \phi^*$.
\begin{equation}
  S = S_{ks} + S_{kp} + S_{s+} + S_{s-} + S_{pg} + S_{p+} +
S_{p-} \;\;\; .
\end{equation}
The result (4.22) can be extended easily to the case of Dirac
fermion using definitions (3.44). In this way we obtain the
result  given in (2.15).

\section{Symmetric form of the effective action }
\setcounter{equation}{0}

\hspace*{4mm}The resulting form of
the effective action (4.22) has still some
unpleasant features.
There is an inverse derivative $\partial_-^{-1}$ in the kinetic
term of the $\chi_-$ field and this term is not symmetric under
$+ \leftrightarrow -$. On the other hand $\chi_-$ and $\chi_+$
are not independent, see (3.46).

The separation of longitudinal and transverse dimensions in the
peripheral high-energy scattering should be reflected by the
form of the action. Separating modes according to kinematics we
have introduced different fields for exchanged and scattering
particles. Each term in (4.22) is Lorentz and scale symmetric.
We would like to have these symmetries separately in the
longitudinal and transverse subspaces. The gluonic terms (3.35),
(3.36) obey this symmetry property, except the kinetic term for
$\phi$, where it holds only after replacing $\Box$ by $4 \partial_+
\partial_-$.

Trying to assign separate longitudinal and transverse scale
dimensions to all fields, we arrive at an unsatisfactory
conclusion for fermions: The supersymmetry parameter $\alpha_+$
carries dimension $- \frac{1}{2}$, and this is clearly a
longitudinal dimension, dim$_\alpha \alpha_+ = \left( \begin{array}{c}
- \frac{1}{2} \\ 0  \end{array} \right) $. From this we have
\begin{eqnarray}
 &&  \mbox{dim}_\alpha \phi = \left( \begin{array}{c} 0 \\ 0 \end{array}
     \right) \; , \quad \mbox{dim}_\alpha {\cal A}_\pm =
      \left( \begin{array}{c} 1 \\ 0  \end{array} \right)
\nonumber \\
 && \mbox{dim}_\alpha \chi_\pm = \left( \begin{array}{c}
     \frac{1}{2} \\ 0  \end{array} \right) \; , \quad \mbox{dim}_\alpha
     a_+ = \left( \begin{array}{c} \frac{3}{2} \\ 0 \end{array} \right)
     \; , \quad \mbox{dim}_\alpha a_- = \left( \begin{array}{c}
     \frac{1}{2} \\ 1 \end{array} \right)
\end{eqnarray}
This assignment works only for those fermionic terms
reconstructed by the $\alpha_+$ supersymmetry (4.3), (4.4), i.e.
$S_{s-}, S_{p-}$. It is asymmetric under $+ \leftrightarrow -$.

However, the kinetic terms of type ${\cal A}$ fields and the
interaction terms describing gluon production are symmetric
under $+ \leftrightarrow -$ and $\phi \leftrightarrow \phi^*$.
These terms allow also a symmetric assignment of scaling dimensions.
\begin{equation}
  \mbox{dim}_\beta \chi_\pm = \left( \begin{array}{c} 0 \\ \frac{1}{2}
     \end{array} \right) \, , \quad \mbox{dim}_\beta a_\pm = \left(
     \begin{array}{c} 1 \\ \frac{1}{2} \end{array} \right)
\end{equation}
The gluon fields have dim$_\alpha = \mbox{dim}_\beta$. We see
that also fermion production terms $S_{p-} + S_{p+}$ involving both
$a_\pm$ and $\chi_\pm$ are compatible with the assignment
dim$_\beta$. It is possible to achieve this also for the
remaining terms by applying the relation (3.46) connecting
$\chi_+$ and $\chi_-$. It is clearly not compatible with
definite transverse and longitudinal dimensions of $\chi_\pm$.
At first glance this procedure seems to be not unique. There are
two ways to rewrite the kinetic and the scattering terms of
$\chi_-$ compatible with (5.2). However, it is reasonable to
include from each of these terms both possible results. The
effective action for gluons and for one Majorana fermion in the
adjoint representation then takes the following form
\begin{eqnarray}
  && S_{ks} = \int d^4x \left\{ - \frac{1}{2} (\partial^* \phi^{a})\Box
     \partial \phi^{a *} + \frac{i}{2} \chi_+^a \Box \partial \chi_-^a +
     \frac{i}{2} \chi_+^{a*} \Box \partial^* \chi^{a*}_- \right\}
         \nonumber \\
  && S_{kp} = \int d^4x \left[ -2A^a_+\partial\partial^* A^a_- -
      ia^a_+\partial a^a_- - ia^{a*}_+\partial^* a^{a*}_- \right]
       \nonumber \\
 && S_{s-} = - g\int d^4x \left\{ \frac{i}{2}[(\partial_-\partial^*\phi)
    T^a(\partial\phi^*)A^{a}_+ +
         (\partial_-\partial\phi^*)T^a(\partial^*\phi) A^a_+]
          \right. \nonumber \\
        && \left. + A^{a}_+
          (\partial_-\chi_+)T^a(\partial\chi_-) +
          A^a_+(\partial^*\chi^*_-)T^a(\partial_-
          \chi^*_+) \right. \nonumber \\
&& \left. - (\partial\phi^{a *})(\partial_-\chi_+)T^aa_+ -
     (\partial^*\phi^a)(\partial_-\chi^*_+)T^aa^*_+ \; \right\} \nonumber\\
&& S_{s+} = - g\int d^4x \left\{ \frac{i}{2}[(\partial_+\partial^*\phi^*)
    T^a(\partial\phi) A^{a}_- +
         (\partial_+\partial\phi)T^a(\partial^*\phi^*) A^a_-]
          \right. \nonumber \\
        && \left. + A^a_-
          (\partial_+\chi_-)T^a(\partial\chi_+) +
          A^a_-(\partial^*\chi^*_+)T^a(\partial_+
          \chi^*_-) \right. \nonumber \\
&& \left. - (\partial\phi^{a})(\partial_+\chi_-)T^aa_- -
     (\partial^*\phi^{a *})(\partial_+\chi^*_-)T^aa^*_- \;
     \right\} \nonumber\\
&&   S_p = g \int d^4 x \left[ \phi^a(\partial A_-)T^a(\partial^*
       A_+) - \phi^{a *}(\partial^*
      A_-)T^a(\partial A_+ )   \right. \nonumber  \\
  &&    - \left. \frac{i}{2}[\phi^a(-a_+T^a(\partial
       a _-) + (\partial^*a^*_+)T^aa^*_-)    \right.
     \nonumber  \\
 &&    + \left. \phi^{a *}(a^*_+T^a(\partial^*a^*_-) -
      (\partial a_+)T^aa_-)]  \right. \nonumber \\
 &&   + \left. i[{\chi}^*_-T^aa^*_-(\partial^* A^a_+) -
    {\chi}_-T^aa_-(\partial  A^a_+) \right.
     \nonumber \\
  &&  + \left. {\chi}^*_+T^aa^*_+ ( \partial^*  A^a_-) -
       {\chi}_+T^aa_+(\partial A^a_-]
             \right]
\end{eqnarray}
The relation (3.46) has to be considered as an additional
constraint.
The symmetry properties in the subspaces hold for the kinetic
term in Eq.(5.3) only after replacing $\Box$ by $4 \partial_+
\partial_-$.  The behaviour of the fields under Lorentz
transformations in longitudinal and rotations in transverse
subspaces is as follows. $\phi$ is scalar in both subspaces.
$A_\pm$ are vector components in longitudinal but scalar
in transverse subspaces. $\chi_\pm$ and $a_\pm$ behave as
two-dimensional spinors in both longitudinal and transverse
subspaces, where the indices $\pm$ refer to the longitudinal
space and the presence of absence of the sign of complex
conjugation $(^*)$ indicates the behaviour under transverse rotations.

The new form of the effective action, in particular the kinetic
term in Eq.(5.3), leads us to a new supersymmetry $\Delta$.
After removing from (5.3) all terms involving $a^*_\pm$ the
remaining expression is invariant under the following
transformation acting on $A_\pm , \phi, \phi^*, \chi_\pm,
\chi^*_\pm, a_\pm $
\begin{eqnarray}
 & \Delta {\cal A}_\pm = \frac{i}{\sqrt{2}} \beta_\pm a_\pm
   & \Delta a_\pm =  \sqrt{2} \beta_\mp \partial^* {\cal A}_\pm
    \nonumber \\
 & \Delta \phi = i\sqrt{2} \beta_- \chi_+ & \Delta \phi^* =
     i\sqrt{2} \beta_+ \chi_- \nonumber \\
 & \Delta \chi_- = \frac{1}{\sqrt{2}} \beta_- \partial^* \phi^*
   & \Delta \chi_+ = \frac{1}{\sqrt{2}} \beta_+ \partial^* \phi
\end{eqnarray}
The transformations of $\chi^*_\pm$ follow from the relation (3.46).
It is convenient first to rewrite all the terms with
$\chi^*_\pm$ by using (3.46) into terms involving $\chi_\pm$ before
checking the invariance of the action (5.3) under
transformations (5.4). \\
The supersymmetry algebra closes on transverse translations.
\begin{eqnarray}
 && \Delta = \beta_- \Delta_+ + \beta_+ \Delta_- \nonumber \\
 && \Delta_+^2 = \Delta_-^2 = 0 \qquad \Delta_+ \Delta_- + \Delta_-
    \Delta_+ = i \partial^*
\end{eqnarray}
There is also the analogous supersymmetry $\Delta^*$ under which
(5.3) is invariant after removing the terms involving $a_\pm$.
The corresponding transformation laws are obtained from (5.4) by
formal complex conjugation.

The new transverse supersymmetry $\Delta$ allows to reconstruct
the terms involving $\chi_\pm, a_\pm$ from the pure gluonic
ones. The terms involving $\chi^*_\pm, a^*_{\pm}$ are reconstructed
using $\Delta^*$ or simply by complex conjugation. Compared to
the original longitudinal supersymmetry the transverse
supersymmetry refers to the final form of the effective action.

In the present form the symmetries discussed above are to be considered
merely as relations between interaction terms. We have to keep in mind
the relation (3.46) between $\chi_+$ and $\chi_-$ and also that the
transverse parts of the kinetic terms violates the separate scaling
symmetries.

We can replace the fermionic kinetic terms in (5.3) which are of second
order in time by terms involving just a first order time derivative.
\begin{eqnarray}
   S_{ks}^{\chi} &=& 4i \int d^4x \left\{ (\partial^*\chi^{a *}_+)
   (\partial_-\partial\chi^a_+) + (\partial^*\chi^{a *}_-)
   (\partial_+\partial\chi^a_-) \right. \nonumber \\
   &+& \left. (\partial^*\chi^{a *}_+)(\partial\partial^*\chi^{a
    *}_-) + (\partial^*\chi^{a *}_-)(\partial\partial^*\chi^{a
   *}_+) \; \right\}
\end{eqnarray}
This is just the representation of the original free Majorana action in
components $\partial^{*} \chi^{*}_{\pm} $ and
 $\partial \chi_{\pm} $. After this replacement the constraint (3.46)
can be dropped. $\chi_+$ and $\chi_-$ are now independent. Together with
their complex conjugates they represent the field variables and
conjugate momenta of a Majorana fermion.

The symmetry properties discussed above do not apply to this form of the
fermion kinetic terms.

\section{The Weizs\"acker-Williams type relations}
\setcounter{equation}{0}
\hspace*{4mm} Our derivation of the effective action is based on
separating modes related to scattering and exchanged particles.
Correspondingly we have obtained vertices describing scattering
of particles and effective vertices describing particle
production. It turns out that between the scattering and
production part of the effective action there are relations
which we call Weizs\"acker-Williams (W-W) type relations, which
to each term in the production part attributes in a definite way a
term in the scattering part.
These symmetry relations are the traces of the common origin of
fields of type A $(\phi, \chi_\pm)$ and of type ${\cal A} (
A_\pm, a_\pm)$ as certain modes of one and the same gluon or
quark field. They hold both for the cases of Dirac and Majorana
fermions.

Let us start with the term in $S_p$ (2.19) describing the gluon
production from the $t$-channel gluonic line
\begin{equation}
g \int d^4x \left[\phi^a(\partial A_-)T^a
(\partial^* A_+) - \phi^{a *}(\partial^* A_-)T^a
(\partial A_+)\right]
\end{equation}
Separately, each term in (6.1) corresponds to a definite
helicity produced gluon.

Let us concentrate for definiteness on the term in (6.1)
involving $\phi$ field. Consider the case, that the field $
A_-$ represents only "soft" modes in the sence that
\begin{equation}
           \partial\partial^* A_-  \rightarrow 0 \;\;.
\end{equation}
Then the integration by part in the considered term of (6.1)
leads to
\begin{equation}
     - g \int d^4x (\partial^*\phi)T^a(\partial A_-)
         A^a_+   \;\;.
\end{equation}
Now we apply the following substitution rule to the "soft" field
$A_-$
\begin{equation}
    A_- \Rightarrow - i\partial_-\phi^*
\end{equation}
which leads to the formula
\begin{equation}
   ig\int d^4x
(\partial^*\phi)T^a(\partial_-\partial\phi^*) A^a_+ \;\;.
\end{equation}
Expression (6.5) coincides with the gluonic term in the
scattering part $S_{s-}$ (2.17) (we use the fact that the
longitudinal momentum component $k_-$ of $A_+$ is small).
The substitution rule (6.4) implies the "softeness" of the field
$\phi^*$ in (6.5). Also, let us note that because $A_-$
field is real one, the "soft" $\phi^*$ field is pure imaginary.

The analogous procedure can be applied in the case when $
A_+$ field in the first term of (6.1) is "soft" one
\begin{equation}
       \partial\partial^* A_+ \rightarrow 0 \;\;.
\end{equation}
The integration by part and use of the substitution rule
\begin{equation}
    A_+ \Rightarrow i\partial_+\phi^*
\end{equation}
leads to
\begin{equation}
     ig \int d^4x A^a_-(\partial\phi)T^a
         (\partial_+\partial^*\phi^*)
\end{equation}
i.e. we recover the gluonic term in the scattering part $S_{s+}$
(2.18), with the $\phi^*$ field being a "soft" one.

The similar considerations performed in the case of the term in
(6.1) involving the $\phi^*$ field lead to the substitution rules
for the "soft" $A_-$ or $A_+$ fields
\begin{eqnarray}
     && A_- \Rightarrow i\partial_-\phi \nonumber \\
         && A_+ \Rightarrow -i\partial_+\phi
\end{eqnarray}
and to the expressions (6.5) and (6.8) involving "soft" $\phi$ field.

The W-W substitution rules (6.4),(6.7),(6.9) can be understood
from the definitions of $\phi$ (3.24) and ${\cal A}_{\pm}$
(3.12),(3.14) in terms of the original gluonic field.

Similar W-W type relations hold between vertices involving
fermionic fields. Let us consider the gluon production from
$t$-channel fermion line (see (2.19))
\begin{eqnarray}
 - \frac{ig}{2}\int d^4x &[&
\phi^a((\partial^*\bar{a}_+)t^aa^*_-) -
\bar{a}^*_+t^a(\partial a_-)) +  \nonumber \\
  &+& \phi^{a *}(\bar{a}_+t^a(\partial^*a^*_-) -
 (\partial\bar{a}^*_+)t^aa_-)] \; \;\;.
 \end{eqnarray}
 We assume that $\bar{a}^*_+$ and $\bar{a}_+$ are "soft" i.e.
 \begin{equation}
 \partial^*\bar{a}_+ \rightarrow 0 \;\;\; \mbox{and} \;\;\; \partial
 \bar{a}^*_+ \rightarrow 0  \;\;.
 \end{equation}
Then the integration by part in (6.10) and use of the
substitution rules
\begin{equation}
\bar{a}^*_+ \Rightarrow -2i\partial^*\bar{\chi}_+ \;\;\;\;\;
\bar{a}_+ \Rightarrow 2i\partial \bar{\chi}^*_+
\end{equation}
lead to expression
\begin{equation}
-g\int d^4x \left\{
(\partial\phi^a)(\partial^*\bar{\chi}_+)t^aa_- +
(\partial^*\phi^{a *})(\partial \bar{\chi}^*_+)t^aa^*_- \; \right\}
\end{equation}
i.e. we recover (see (3.46)) the two last vertices in $S_{s
+}$ (2.18). \\
If we assume that in the expression (6.10) $a_-$ and $a^*_-$ are
the "soft" fields i.e.
\begin{equation}
\partial a_- \rightarrow 0 \;\;\;\;\; \partial^*a^*_-
\rightarrow 0
\end{equation}
then with the use of the substitution rules
\begin{equation}
a^*_- \Rightarrow 2i\partial\chi_- \;\;\;\; a_- \Rightarrow
-2i\partial^* \chi^*_-
\end{equation}
and relations (3.46),(3.47) we recover the vertices in the third
line of $S_{s -}$ (2.17).

As a final example let us consider the following part of the
fermion production vertices in $S_p$ (2.19)
\begin{equation}
-ig\int d^4x \left\{\bar{\chi}^*_-t^aa_-(\partial A^a_+) -
\bar{\chi}_-t^aa^*_-(\partial^* A^a_+) \; \right\} \;.
\end{equation}
Restricting $a_-$, $a^*_-$ fields to the "soft" modes (see
(6.14)) and using the rules (6.15) we obtain from (6.16) the
expression
\begin{eqnarray}
&\mbox{}&2g\int d^4x
A^a_+[(\partial\bar{\chi}^*_-)t^a(\partial^* \chi^*_-) +
(\partial^*\bar{\chi}_-)t^a(\partial\chi_-)] \nonumber \\
&=& - 2g\int d^4x A^a_+
[(\partial_-\bar{\chi}_+)t^a(\partial^*\chi^*_-) +
(\partial_-\bar{\chi}^*_+)t^a(\partial\chi_-)]
\end{eqnarray}
i.e. we recover the second line of $S_{s -}$ (2.17). \\
Now, let us assume that in the formula (6.16) the field $A_+$
 is a "soft" one (see (6.6)). Then the substitution rule (6.7)
(see also (6.9)) supplemented by the integration by part leads to
\begin{equation}
g\int d^4x \left\{
(\partial\phi^a)(\partial_+\bar{\chi}^*_-)t^aa_- +
(\partial^*\phi^{a *})(\partial_+\bar{\chi}_-)t^aa^*_- \; \right\}
\end{equation}
coinciding with the last line of $S_{s +}$ (2.18).

Also in the fermionic case the W-W substitution rules can be
understood from the definitions of $\chi_\pm$ and $a_\pm$
(3.44), (3.46) in terms of the original quark field.

\section{Discussion}
\setcounter{equation}{0}
\hspace*{4mm} The effective action (2.15) describes QCD scattering processes
in the multi-Regge regime. It summarizes results about the leading
contribution in this region obtained by analyzing graphs using ideas of
unitarity, $s$-channel helicity conservation and the separation of
longitudinal and transverse dimensions. We have derived the effective action
from QCD by separating the modes of the original gluon and quark fields
according to the multi-Regge kinematics and integrating over modes which do
not correspond to scattering or exchanged particles in the considered
peripheral process. Supersymmetry has been used to obtain the fermionic terms
of the effective action from the pure gluonic ones. This supersymmetry is
the subgroup of the one of the supersymmetric Yang-Mills theory, which is
associated with the direction of the momentum of a incoming particle.
A supersymmetry related to the transverse directions transforms the fermionic
and gluonic terms of the final form of this action into each other. The
vertices corresponding to scattering and production of particles
 are connected to
each other by Weizs\"acker-Williams type relations.

The resulting action reproduces the leading contribution in QCD to high
energy peripheral scattering amplitudes in a most economic way, because
most of the neglegible contributions in this kinematical region are
excluded. Therefore we have a simple and symmetric structure of the result.
The fields describing the scattered and the exchanged gluons and quarks
do not have 4-dimensional Lorentz or Dirac indices any more. They are
attributed a definite behaviour under scaling and Lorentz or rotation
transformations separately in the longitudinal and transverse subspaces,
reflecting the kinematics of the processes where the action applies. The
action is symmetric ( with a modification in the kinetic term of scattering
particles ) under scaling and Lorentz transformations separately in both
subspaces.

The applicability of the multi-Regge effective action is of
course restricted to the multi-Regge kinematics in the
scattering amplitudes, a condition which has been used in all
steps of the derivation. In this regime the action (2.15)
reproduces the leading contribution correctly. This is because
in the considered asymptotics intermediate states in all
subenergy channels obeying multi-Regge kinematics dominate.
Contributions from pairs of particles in non-multi-Regge
configuration or from loops with particles far off shell lead to
corrections to our result \cite{FadLip}.

In the dominating kinematics the longitudinal momenta of the
exchanged particles are small compared to the their transverse
momenta. This condition is not included explicitly in the
propagators of $A_{\pm}$ and it may be necessary in some
loop integrals to impose it by a cut-off. A modification of the
effective vertices would be desirable, which suppresses
contributions from outside of the multi-Regge region.

In the $s$-channel intermediate states the scattering particles
can be virtual as long as their momenta squared are small
compared to the subenergies. One may fix an upper limit of the
virtualness of scattering particles. Then additional triple
vertices involving exchange particles (type ${\cal A}$) have to
be included into the action.

In the approach by Verlinde \cite{Verlinde} a scaling argument leads to
neglecting completely the contribution of the transverse field strength
components to the gauge field action.  Different to
Verlinde the contributions  from the transverse field
strength squared are essential for our result. The  integration over
$A'_+$ produces the interaction terms from which we obtain finally the
particle production vertex. It seems that the approach
\cite{Verlinde} does not apply to the inelastic amplitudes.

Verlinde established a remarkable connection of their effective action to
the two-dimensional WZNW model. It was shown earlier \cite{quasielast}
that constructing amplitudes obeying just elastic unitarity in $s$- and
$u$- channels results in an factorizable $S$-matrix coinciding with the
results for $\sigma$-models \cite{Zamol}.

The multi-Regge effective action is the first step in a new approach
to high-energy peripheral scattering overcoming the deficits of both
the eikonal and the leading logarithmic approximations. With this action we
would like to calculate QCD amplitudes including the effects of multi-particle
intermediate states and satisfying the unitarity conditions in all
sub-energy channels. The Weizs\"acker-Williams type relations support
the expectation that also the $t$-channel unitarity conditions can be
obeyed in this approach.

The separation of longitudinal and transverse subspaces and of scattering
and exchanged fields suggests to apply functional methods for further
simplifications. The exchanged fields $(A_{\pm}, a_{\pm})$ are closely
related to reggeons. One should try to find a tractable representation of the
effective reggeon field theory. Conformal symmetry in the two-dimensional
transverse space is expected to play an important role.  \\

\newpage
\noindent {\Large\bf Appendix}
\renewcommand{\theequation}{A.\arabic{equation}}

\vspace{1cm}
Below we summarize our notation  and conventions related to the
fermionic sector of the effective action.

We introduce four nilpotent matrices $\gamma_\pm$ and $\gamma,
\gamma^*$ according to the formulas:
\begin{equation}
 \gamma_\pm \equiv \gamma_0 \pm \gamma_3 \qquad \gamma \equiv
  \gamma^1 + i \gamma^2 \qquad \gamma^* \equiv \gamma^1 - i \gamma^2
\end{equation}
They satisfy the following relations
\begin{eqnarray}
  \{ \gamma_+, \gamma_- \} = 4  && \{ \gamma, \gamma^* \} = -4
        \nonumber \\
  ( \gamma_\pm)^\dagger = \gamma_\mp  && (\gamma)^\dagger = - \gamma^*
\end{eqnarray}
{}From the matrices (A.1) we built four projection operators
\begin{eqnarray}
  & & \Pi_+ = \frac{1}{4} \gamma_- \gamma_+  \qquad P_+ = - \frac{1}{4}
          \gamma \gamma^* \nonumber \\
 &  & \Pi_- = \frac{1}{4} \gamma_+ \gamma_- \qquad P_- = - \frac{1}{4}
          \gamma^* \gamma
\end{eqnarray}
which lead to the decomposition of the unit operator in the
spinor space
 \begin{equation}
  {\bf 1} = \Pi_+ P_+ + \Pi_+ P_- + \Pi_- P_+ +
       \Pi_- P_-
\end{equation}
Each term on the r.h.s. of Eq. (A.4) defines the corresponding
fermionic subspace on which it projects. We choose in each of
these subspaces one basic spinor $u_{ij} (i,j = + \; \; \mbox{or}
\;\; -)$ satisfying
\begin{equation}
  \Pi_i P_j u_{ij} = u_{ij}
\end{equation}
and being the Majorana one
\begin{equation}
  u_{i+} = C \bar{u}_{i-}^{T}
\end{equation}
For the basic spinors $u_{ij}$ we assume \newline
a) the normalization conditions
\begin{eqnarray}
  &&  \bar{u}_{+j} \gamma_{+} u_{+j} = 1 \nonumber \\
  &&  \bar{u}_{-j} \gamma_{-} u_{-j} = 1 \qquad j = +, -
\end{eqnarray}
b) the phase conventions
\begin{eqnarray}
  u_{++}  = - \frac{1}{2} \gamma_- u_{-+} &&
  u_{++} =  \frac{1}{2} \gamma u_{+-} \nonumber \\
  u_{+-} = \frac{1}{2} \gamma_- u_{--} &&
    u_{+-} = - \frac{1}{2} \gamma^* u_{++}
\end{eqnarray}
All other matrix elements between the basic spinors
$u_{ij}$ can be calculated with the help of Eqs.~(A.6), (A.7) and (A.8).
With all these ingredients the arbitrary spinor can be
decomposed in the basis $u_{ij}$ and this fact leads in
particular to Eq.~(3.20).

\vspace{2cm}

\noindent {\Large\bf Figure Captions}

\vspace{1cm}
\begin{tabular}{ll}
 Fig. 1 & The general inelastic process in the multi-regge
            kinematics \\
 Fig. 2 & The gluon production in the gluon-fermion scattering \\
 Fig. 3 & The gluon production in the gluon-gluon scattering \\
 Fig. 4 & The fermion production in the gluon-fermion scattering
          \\
 Fig. 5 & Vertices resulting from the second term in the action ~(3.4) \\
        &  by separating fields modes of types ${\cal A}$(dashed line), \\
        &  A(full line) and A$_1$( double line). We imagine the graphs \\
        &  as parts of graph for the multiparticle amplitude in Fig. 1. \\
        &  The $s$-channel goes vertically and the $t$-channel
          horizontally. \\
        &  Lines going more to the right carry smaller $k_-$. \\
 Fig. 6 & Diagrams contributing to the effective production vertex.\\
 Fig. 7 & Contribution from the third term of the action (3.4). The \\
        &  blob represents the non-local vertex involving $\partial^{-2}_-$.

 \end{tabular}
\newpage


\setcounter{page}{32}
\noindent {\Large\bf Table 1}
\begin{center}
  {\Large\bf Feynman rules for the effective action}
\end{center}

\vspace{1cm}
\noindent We introduce the notation
\[  {\cal F}\left[ f(\mbox{k}) \right] =
     \int \frac{d^4 \mbox{k}}{(2\pi)^4} e^{-i\mbox{k}(x-y)}
     f(\mbox{k})  \]

\noindent The propagators of $t$-channel fields:
\[ \langle 0|T ({\cal A}^a_+(x) {\cal A}^b_-(y))|0\rangle =
   {\cal F}\left[\frac{(2i \delta^{ab})}{kk^*} \right] \]
\[
  \langle 0|T(a^a_+ (x) \bar{a}^{b*}_-(y))|0\rangle
  = \langle 0|T(a^a_- (x)
   \bar{a}^{b*}_+ (y))|0\rangle
   = {\cal F}\left[
   \frac{(2i \delta^{ab})}{k^*} \right]  \]
 \[ \langle 0|T (a^{a*}_+ (x) \bar{a}^b_- (y))|0 \rangle
 = \langle 0|T (a^{a*}_-
  (x) \bar{a}^b_+ (y))|0 \rangle
  = {\cal F}\left[
  \frac{(2i \delta^{ab})}{k} \right]   \]
The propagators of $s$-channel fields:
\[ \langle 0|T (\phi^a(x) \phi^{b*} (y)) |0 \rangle =
  {\cal F}\left[
  \frac{8i \delta^{ab}}{(\mbox{k}^2 + i\varepsilon) kk^*}\right] \]

\[ \langle 0|T(\chi^a_- (x) \bar{\chi}^b_- (y))|0\rangle =
{\cal F}\left[
   \frac{2i \delta^{ab}k_-}{(\mbox{k}^2 + i\varepsilon)kk^*}\right]  \]

\[  \langle 0|T(\chi^{a*}_-(x) \bar{\chi}^{b*}_- (y)) |0 \rangle =
   {\cal F}\left[
   \frac{2i\delta^{ab}k_-}{(\mbox{k}^2 + i\varepsilon) kk^*} \right]
   \;\;\;\;\mbox{etc.}   \]

 Below we present the interaction vertices. The $s$-channel
momenta flow from the bottom of the page to the top, whereas the
$t$-channel momenta flow from the right to the left of the page.
The signs + or - denote the corresponding particle helicities.
\newpage

\setlength{\textheight}{24cm}
\setlength{\textwidth}{20cm}
\setlength{\oddsidemargin}{0.0mm}
\setlength{\evensidemargin}{0.0mm}
\pagestyle{empty}
\footheight0.0pt
\vspace*{-3cm}
\input FEYNMAN


\begin{picture}(51000,69000)

\drawline\gluon[\N\CURLY](3000,63000)[3]
\drawline\gluon[\S\FLIPPEDCURLY](3000,63000)[3]
\drawline\gluon[\E\CURLY](3000,63000)[4]
\put(2000,66000){+}
\put(4000,66000){p$'$}
\put(2000,60000){+}
\put(4000,60000){p}
\put(6000,61000){q}
\put(7000,63000){$\frac{ig}{2}(p'_- +
p_-)e_R(p')T^ae_R(p)^*{\cal A}^a_+$}

\drawline\gluon[\N\CURLY](3000,55500)[3]
\drawline\gluon[\S\FLIPPEDCURLY](3000,55500)[3]
\drawline\gluon[\E\CURLY](3000,55500)[4]
\put(2000,58000){$-$}
\put(2000,52500){$-$}
\put(7000,55500){$\frac{ig}{2}(p'_- +
p_-)e_R(p')^*T^ae_R(p){\cal A}^a_+$}

\drawline\fermion[\N\REG](3000,48000)[3000]
\drawarrow[\N\ATTIP](\pmidx,\pmidy)
\drawline\fermion[\S\REG](3000,48000)[3000]
\drawarrow[\N\ATTIP](\pmidx,\pmidy)
\drawline\gluon[\E\CURLY](3000,48000)[4]
\put(2000,50000){+}
\put(2000,45000){+}
\put(7000,48000){$\frac{ig}{2}(p_- +
p'_-)\chi_R(p')t^a\chi_R(p)^*{\cal A}^a_+$}

\drawline\fermion[\N\REG](3000,40500)[3000]
\drawarrow[\N\ATTIP](\pmidx,\pmidy)
\drawline\fermion[\S\REG](3000,40500)[3000]
\drawarrow[\N\ATTIP](\pmidx,\pmidy)
\drawline\gluon[\E\CURLY](3000,40500)[4]
\put(2000,43000){$-$}
\put(2000,37500){$-$}
\put(7000,40500){$\frac{ig}{2}(p_- +
    p'_-)\chi_R(p')^*t^a\chi_R(p){\cal A}^a_+$}

\drawline\fermion[\N\REG](3000,33000)[3000]
\drawarrow[\N\ATTIP](\pmidx,\pmidy)
\put(2000,35500){+}
\drawline\fermion[\E\REG](3000,33000)[3000]
\drawarrow[\W\ATTIP](\pmidx,\pmidy)
\drawline\gluon[\S\FLIPPEDCURLY](3000,33000)[3]
\put(2000,30000){+}
\put(7500,33000){$ ig\sqrt{p'_-}\chi_R(p')t^aa_+e^a_R(p)^*$}

\drawline\fermion[\N\REG](3000,25500)[3000]
\drawarrow[\N\ATTIP](\pmidx,\pmidy)
\put(2000,28000){$-$}
\drawline\fermion[\E\REG](3000,25500)[3000]
\drawarrow[\W\ATTIP](\pmidx,\pmidy)
\drawline\gluon[\S\FLIPPEDCURLY](3000,25500)[3]
\put(2000,23000){$-$}
\put(7500,25500){$ ig\sqrt{p'_-}\chi_R(p')^*t^aa_+^*e^a_R(p)$}

\drawline\gluon[\N\CURLY](3000,18000)[3]
\put(2000,21000){+}
\drawline\fermion[\S\REG](3000,18000)[3000]
\drawarrow[\N\ATTIP](\pmidx,\pmidy)
\put(2000,15500){+}
\drawline\fermion[\E\REG](3000,18000)[3000]
\drawarrow[\E\ATTIP](\pmidx,\pmidy)
\put(7500,18000){$ ige^a_R(p')a^*_+t^a\chi_R(p)^*\sqrt{p_-}$}

\drawline\gluon[\N\CURLY](3000,10500)[3]
\put(2000,13000){$-$}
\drawline\fermion[\S\REG](3000,10500)[3000]
\drawarrow[\N\ATTIP](\pmidx,\pmidy)
\put(2000,8000){$-$}
\drawline\fermion[\E\REG](3000,10500)[3000]
\drawarrow[\E\ATTIP](\pmidx,\pmidy)
\put(7500,10500){$ ige^a_R(p')^*a_+t^a\chi_R(p)\sqrt{p_-}$}


\drawline\gluon[\N\CURLY](30000,63000)[3]
\drawline\gluon[\S\FLIPPEDCURLY](30000,63000)[3]
\drawline\gluon[\W\CURLY](30000,63000)[4]
\put(31000,66000){+}
\put(29000,66000){p$'$}
\put(31000,60000){+}
\put(29000,60000){p}
\put(27000,62000){q}
\put(31500,63000){$\frac{ig}{2}(p'_+ +
p_+)e_L(p')^*T^ae_L(p){\cal A}^a_-$}

\drawline\gluon[\N\CURLY](30000,55500)[3]
\drawline\gluon[\S\FLIPPEDCURLY](30000,55500)[3]
\drawline\gluon[\W\CURLY](30000,55500)[4]
\put(31000,58000){$-$}
\put(31000,52500){$-$}
\put(31500,55500){$\frac{ig}{2}(p_+ +
p'_+)e_L(p')T^ae_L(p)^*{\cal A}^a_-$}

\drawline\fermion[\N\REG](30000,48000)[3000]
\drawarrow[\N\ATTIP](\pmidx,\pmidy)
\drawline\fermion[\S\REG](30000,48000)[3000]
\drawarrow[\N\ATTIP](\pmidx,\pmidy)
\drawline\gluon[\W\CURLY](30000,48000)[4]
\put(31000,50000){+}
\put(31000,45000){+}
\put(32000,48000){$\frac{ig}{2}(p_+ +
p'_+)\chi_L(p')^*t^a\chi_L(p){\cal A}^a_-$}

\drawline\fermion[\N\REG](30000,40500)[3000]
\drawarrow[\N\ATTIP](\pmidx,\pmidy)
\drawline\fermion[\S\REG](30000,40500)[3000]
\drawarrow[\N\ATTIP](\pmidx,\pmidy)
\drawline\gluon[\W\CURLY](30000,40500)[4]
\put(31000,43000){$-$}
\put(31000,37500){$-$}
\put(32000,40500){$\frac{ig}{2}(p_+ +
p'_+)\chi_L(p')t^a\chi_L(p)^*{\cal A}^a_-$}

\drawline\fermion[\S\REG](30000,33000)[3000]
\drawarrow[\N\ATTIP](\pmidx,\pmidy)
\put(31000,35500){+}
\drawline\fermion[\W\REG](30000,33000)[3000]
\drawarrow[\W\ATTIP](\pmidx,\pmidy)
\drawline\gluon[\N\FLIPPEDCURLY](30000,33000)[3]
\put(31000,30000){+}
\put(32000,33000){$- ige^a_L(p')^*a_-t^a\chi_L(p)\sqrt{p_+}$}

\drawline\fermion[\S\REG](30000,25500)[3000]
\drawarrow[\N\ATTIP](\pmidx,\pmidy)
\put(31000,28000){$-$}
\drawline\fermion[\W\REG](30000,25500)[3000]
\drawarrow[\W\ATTIP](\pmidx,\pmidy)
\drawline\gluon[\N\FLIPPEDCURLY](30000,25500)[3]
\put(31000,23000){$-$}
\put(32000,25500){$- ige^a_L(p')a^*_-t^a\chi_L(p)^*\sqrt{p_+}$}

\drawline\gluon[\S\CURLY](30000,18000)[3]
\put(31000,21000){+}
\drawline\fermion[\N\REG](30000,18000)[3000]
\drawarrow[\N\ATTIP](\pmidx,\pmidy)
\put(31000,15500){+}
\drawline\fermion[\W\REG](30000,18000)[3000]
\drawarrow[\E\ATTIP](\pmidx,\pmidy)
\put(32000,18000){$- ig\sqrt{p'_+}\chi_L(p')^*t^aa^*_-e^a_L(p)$}

\drawline\gluon[\S\CURLY](30000,10500)[3]
\put(31000,13000){$-$}
\drawline\fermion[\N\REG](30000,10500)[3000]
\drawarrow[\N\ATTIP](\pmidx,\pmidy)
\put(31000,8000){$-$}
\drawline\fermion[\W\REG](30000,10500)[3000]
\drawarrow[\E\ATTIP](\pmidx,\pmidy)
\put(32000,10500){$- ig\sqrt{p'_+}\chi_L(p')t^aa_-e^a_L(p)^*$}

\put(10000,3000){Table 1 : the scattering vertices}
\put(24000,100){33}

 \end{picture}
\newpage

\vspace*{-3cm}


\begin{picture}(51000,69000)

\drawline\gluon[\N\CURLY](9000,63000)[3]
\drawarrow[\N\ATTIP](\pmidx,\pmidy)
\put(8000,66000){+}
\put(9500,66000){k}
\drawline\gluon[\W\FLIPPEDCURLY](9000,63000)[6]
\drawarrow[\W\ATTIP](\pmidx,\pmidy)
\put(3500,64000){q$_1$}
\drawline\gluon[\E\CURLY](9000,63000)[6]
\drawarrow[\W\ATTIP](\pmidx,\pmidy)
\put(14000,64000){q$_2$}
\put(100,60000){$-\frac{ig}{\sqrt{2}}(q^*_1{\cal A}_-(q_1))T^a(q_2
 {\cal A}_+(q_2))\frac{e^a_L(k)^*}{k^*}$}

\drawline\gluon[\N\CURLY](9000,54000)[3]
\put(8000,57000){+}
\drawline\fermion[\W\REG](9000,54000)[6000]
\drawarrow[\W\ATTIP](\pmidx,\pmidy)
\put(3500,55000){+}
\drawline\fermion[\E\REG](9000,54000)[6000]
\drawarrow[\W\ATTIP](\pmidx,\pmidy)
\put(14500,55000){+}
\put(1000,51000){$\frac{ig}{\sqrt{2}}(q^*_1a_-(q_1))t^aa_+(q_2)
\frac{e^a_L(k)^*}{k^*}$}

\drawline\gluon[\N\CURLY](9000,45000)[3]
\put(8000,48000){+}
\drawline\fermion[\W\REG](9000,45000)[6000]
\drawarrow[\W\ATTIP](\pmidx,\pmidy)
\put(3500,46000){$-$}
\drawline\fermion[\E\REG](9000,45000)[6000]
\drawarrow[\W\ATTIP](\pmidx,\pmidy)
\put(14500,46000){$-$}
\put(1000,42000){$\frac{ig}{\sqrt{2}}a^*_-(q_1)t^a(q_2a^*_+(q_2))
\frac{e^a_L(k)^*}{k^*}$
}

\drawline\gluon[\N\CURLY](9000,36000)[3]
\put(8000,39000){+}
\drawline\fermion[\W\REG](9000,36000)[6000]
\drawarrow[\E\ATTIP](\pmidx,\pmidy)
\put(3500,37000){+}
\drawline\fermion[\E\REG](9000,36000)[6000]
\drawarrow[\E\ATTIP](\pmidx,\pmidy)
\put(14500,37000){+}
\put(1000,33000){$\frac{ig}{\sqrt{2}}(q_2a^*_+(q_2))t^aa^*_-(q_1)
\frac{e^a_L(k)^*}{k^*}
$}

\drawline\gluon[\N\CURLY](9000,27000)[3]
\put(8000,30000){+}
\drawline\fermion[\W\REG](9000,27000)[6000]
\drawarrow[\E\ATTIP](\pmidx,\pmidy)
\put(3500,28000){$-$}
\drawline\fermion[\E\REG](9000,27000)[6000]
\drawarrow[\E\ATTIP](\pmidx,\pmidy)
\put(14500,28000){$-$}
\put(1000,24000){$\frac{ig}{\sqrt{2}}a_+(q_2)t^a(q^*_1a_-(q_1))
\frac{e^a_L(k)^*}{k^*}$}

\drawline\fermion[\N\REG](9000,18000)[3000]
\drawarrow[\N\ATTIP](\pmidx,\pmidy)
\put(8000,20500){+}
\drawline\fermion[\W\REG](9000,18000)[6000]
\drawarrow[\E\ATTIP](\pmidx,\pmidy)
\put(3500,19000){+}
\drawline\gluon[\E\FLIPPEDCURLY](9000,18000)[6]
\put(1000,15000){$-\frac{ig}{\sqrt{2}}(\sqrt{k_-}\frac{\chi_R(k)}{k})t^a
a^*_-(q_1)(q_2{\cal A}^a_+(q_2))$}

\drawline\fermion[\N\REG](9000,9000)[3000]
\drawarrow[\N\ATTIP](\pmidx,\pmidy)
\put(8000,11500){+}
\drawline\fermion[\E\REG](9000,9000)[6000]
\drawarrow[\W\ATTIP](\pmidx,\pmidy)
\put(14500,10000){+}
\drawline\gluon[\W\CURLY](9000,9000)[6]
\put(1000,6000){$-\frac{ig}{\sqrt{2}}(q^*_1{\cal A}^a_-(q_1))
(\sqrt{k_+}\frac{\chi_L(k)^*}{k^*})t^aa_+(q_2)$}


\drawline\gluon[\N\CURLY](33000,63000)[3]
\put(32000,66000){$-$}
\drawline\gluon[\W\FLIPPEDCURLY](33000,63000)[6]
\drawline\gluon[\E\CURLY](33000,63000)[6]
\put(24100,60000){$-\frac{ig}{\sqrt{2}}(q_1{\cal A}_-(q_1))T^a
(q^*_2{\cal A}_+(q_2))\frac{e^a_L(k)}{k}$}

\drawline\gluon[\N\CURLY](33000,54000)[3]
\put(32000,57000){$-$}
\drawline\fermion[\W\REG](33000,54000)[6000]
\drawarrow[\W\ATTIP](\pmidx,\pmidy)
\put(27500,55000){+}
\drawline\fermion[\E\REG](33000,54000)[6000]
\drawarrow[\W\ATTIP](\pmidx,\pmidy)
\put(38500,55000){+}
\put(25000,51000){$\frac{ig}{\sqrt{2}}a_-(q_1)t^a
(q^*_2a_+(q_2))\frac{e^a_L(k)}{k}$}

\drawline\gluon[\N\CURLY](33000,45000)[3]
\put(32000,48000){$-$}
\drawline\fermion[\W\REG](33000,45000)[6000]
\drawarrow[\W\ATTIP](\pmidx,\pmidy)
\put(27500,46000){$-$}
\drawline\fermion[\E\REG](33000,45000)[6000]
\drawarrow[\W\ATTIP](\pmidx,\pmidy)
\put(38500,46000){$-$}
\put(25000,42000){$\frac{ig}{\sqrt{2}}(q_1a^*_-(q_1))t^aa^*_+(q_2)
\frac{e^a_L(k)}{k}$}

\drawline\gluon[\N\CURLY](33000,36000)[3]
\put(32000,39000){$-$}
\drawline\fermion[\W\REG](33000,36000)[6000]
\drawarrow[\E\ATTIP](\pmidx,\pmidy)
\put(27500,37000){+}
\drawline\fermion[\E\REG](33000,36000)[6000]
\drawarrow[\E\ATTIP](\pmidx,\pmidy)
\put(38500,37000){+}
\put(25000,33000){$\frac{ig}{\sqrt{2}}a^*_+(q_2)t^a(q_1a^*_-(q_1))
\frac{e^a_L(k)}{k}$}

\drawline\gluon[\N\CURLY](33000,27000)[3]
\put(32000,30000){+}
\drawline\fermion[\W\REG](33000,27000)[6000]
\drawarrow[\E\ATTIP](\pmidx,\pmidy)
\put(27500,28000){$-$}
\drawline\fermion[\E\REG](33000,27000)[6000]
\drawarrow[\E\ATTIP](\pmidx,\pmidy)
\put(38500,28000){$-$}
\put(25000,24000){$\frac{ig}{\sqrt{2}}(q^*_2a_+(q_2))t^aa_-(q_1)
\frac{e^a_L(k)}{k}$}

\drawline\fermion[\N\REG](33000,18000)[3000]
\drawarrow[\N\ATTIP](\pmidx,\pmidy)
\put(32000,20500){$-$}
\drawline\fermion[\W\REG](33000,18000)[6000]
\drawarrow[\E\ATTIP](\pmidx,\pmidy)
\put(27500,19000){$-$}
\drawline\gluon[\E\FLIPPEDCURLY](33000,18000)[6]
\put(25000,15000){$-\frac{ig}{\sqrt{2}}(\sqrt{k_-}\frac{\chi_R(k)^*}{k^*})
t^aa_-(q_1)(q^*_2{\cal A}^a_+(q_2))$}

\drawline\fermion[\N\REG](33000,9000)[3000]
\drawarrow[\N\ATTIP](\pmidx,\pmidy)
\put(33500,11500){$-$}
\drawline\fermion[\E\REG](33000,9000)[6000]
\drawarrow[\W\ATTIP](\pmidx,\pmidy)
\put(38000,10000){$-$}
\drawline\gluon[\W\CURLY](33000,9000)[6]
\put(25000,6000){$-\frac{ig}{\sqrt{2}}(q_1{\cal A}^a_-(q_1))
(\sqrt{k_+}\frac{\chi_L(k)}{k})t^aa^*_+(q_2)$}

\put(12000,2000){Table 1: the production vertices}
\put(24000,10){34}

 \end{picture}


\newpage


\vspace*{-2cm}
\setlength{\textheight}{24cm}
\setlength{\evensidemargin}{0.0mm}
\setlength{\oddsidemargin}{0.0mm}
\pagestyle{empty}
\footheight0.0pt
\textwidth20cm


\begin{picture}(24000,20000)
\drawline\fermion[\N\REG](21000,6000)[6000]
\put(21000,4500){$\mbox{p}_{B}$}
\drawarrow[\N\ATTIP](\pmidx,\pmidy)
\drawline\fermion[\N\REG](21000,12000)[6000]
\drawarrow[\N\ATTIP](21000,16500)
\put(21000,19500){$\mbox{p}_{B'}$}
\drawline\fermion[\N\REG](18000,13500)[4500]
\drawarrow[\N\ATTIP](18000,16500)
\put(18000,19500){$\mbox{k}_{n}$}
\drawline\fermion[\N\REG](15000,13500)[4500]
\drawarrow[\N\ATTIP](15000,16500)
\put(15000,19500){$\mbox{k}_{n-1}$}
\drawline\fermion[\N\REG](9000,13500)[4500]
\drawarrow[\N\ATTIP](9000,16500)
\put(9000,19500){$\mbox{k}_{2}$}
\drawline\fermion[\N\REG](6000,13500)[4500]
\drawarrow[\N\ATTIP](6000,16500)
\put(6000,19500){$\mbox{k}_{1}$}
\drawline\fermion[\N\REG](3000,12000)[6000]
\drawarrow[\N\ATTIP](3000,16500)
\put(3000,19500){$\mbox{p}_{A'}$}
\drawline\fermion[\N\REG](3000,6000)[6000]
\drawarrow[\N\ATTIP](3000,9000)
\put(3000,4500){$\mbox{p}_{A}$}
\drawline\fermion[\W\REG](21000,12000)[18000]
\drawarrow[\W\ATTIP](4500,12000)
\drawarrow[\W\ATTIP](7500,12000)
\drawarrow[\W\ATTIP](16500,12000)
\drawarrow[\W\ATTIP](19500,12000)
\put(4300,12700){$\footnotesize q_1$}
\put(7300,12700){$\footnotesize q_2$}
\put(16300,12700){$\footnotesize q_n$}
\put(18600,12700){$\footnotesize q_{n+1}$}
\put(12000,12000){\oval(18000,3000)}
\put(11000,16500){$.\;\; . \;\; .$}
\put(10500,1500){Fig. 1}
\end{picture}

\vspace*{1cm}

\begin{picture}(24000,21000)
\drawline\gluon[\N\FLIPPED](21000,12000)[5]
\drawarrow[\N\ATTIP](\pmidx,\pmidy)
\put(22000,\pbacky){$\mbox{p}_{B'}$}
\drawline\gluon[\N\FLIPPED](12000,12000)[5]
\drawarrow[\N\ATTIP](\pmidx,\pmidy)
\put(13000,\gluonbacky){k}
\drawline\gluon[\S\FLIPPED](3000,12000)[5]
\drawarrow[\N\ATTIP](\pmidx,\pmidy)
\put(500,\gluonbacky){$\mbox{p}_A$}
\drawline\fermion[\S\REG](21000,12000)[-\gluonlengthy]
\drawarrow[\N\ATTIP](\pmidx,\pmidy)
\put(22000,\fermionbacky){$\mbox{p}_B$}
\drawline\fermion[\W\REG](21000,12000)[9000]
\drawarrow[\W\ATTIP](\pmidx,\pmidy)
\put(\pmidx,13000){$\mbox{q}_2$}
\drawline\fermion[\W\REG](12000,12000)[9000]
\drawarrow[\W\ATTIP](\pmidx,\pmidy)
\put(\pmidx,13000){$\mbox{q}_1$}
\drawline\fermion[\N\REG](3000,12000)[-\gluonlengthy]
\drawarrow[\N\ATTIP](\pmidx,\pmidy)
\put(500,\pbacky){$\mbox{p}_{A'}$}
\put(21000,12000){\circle*{1000}}
\put(3000,12000){\circle*{1000}}
\put(12000,12000){\circle*{1000}}
\put(11000,2500){Fig. 2}
\end{picture}

\vspace*{1cm}

\begin{picture}(24000,18000)
\drawline\fermion[\N\REG](21000,6000)[6000]
\drawarrow[\N\ATTIP](21000,9000)
\put(22000,\fermionfronty){$\mbox{p}_B$}
\drawline\fermion[\N\REG](21000,12000)[6000]
\drawarrow[\N\ATTIP](21000,15000)
\put(22000,\fermionbacky){$\mbox{p}_{B'}$}
\drawline\fermion[\N\REG](3000,6000)[6000]
\drawarrow[\N\ATTIP](3000,9000)
\put(500,\fermionfronty){$\mbox{p}_A$}
\drawline\fermion[\N\REG](3000,12000)[6000]
\drawarrow[\N\ATTIP](3000,15000)
\put(500,\fermionbacky){$\mbox{p}_{A'}$}
\drawline\gluon[\N\FLIPPED](12000,12000)[6]
\drawarrow[\N\ATTIP](\pmidx,\pmidy)
\put(13000,\gluonbacky){k}
\drawline\gluon[\W\FLIPPED](12000,12000)[8]
\drawarrow[\W\ATTIP](\pmidx,\pmidy)
\put(\pmidx,13000){$\footnotesize q_1$}
\drawline\gluon[\E\REG](12000,12000)[8]
\drawarrow[\W\ATTIP](\pmidx,\pmidy)
\put(\pmidx,13000){$\footnotesize q_2$}
\put(3000,12000){\circle*{1000}}
\put(12000,12000){\circle*{1000}}
\put(21000,12000){\circle*{1000}}
\put(11000,2500){Fig. 3}
\put(24000,-2000){35}
\end{picture}

\newpage


\begin{picture}(24000,18000)
\drawline\gluon[\N\FLIPPED](21000,12000)[5]
\drawarrow[\N\ATTIP](\pmidx,\pmidy)
\put(22000,\pbacky){$\mbox{p}_{B'}$ }
\drawline\fermion[\N\REG](12000,12000)[\gluonlengthy]
\drawarrow[\N\ATTIP](\pmidx,\pmidy)
\put(13000,\fermionbacky){k}
\drawline\gluon[\S\FLIPPED](3000,12000)[5]
\drawarrow[\N\ATTIP](\pmidx,\pmidy)
\put(500,\gluonbacky){$\mbox{p}_A$}
\drawline\fermion[\S\REG](21000,12000)[-\gluonlengthy]
\drawarrow[\N\ATTIP](\pmidx,\pmidy)
\put(22000,\fermionbacky){$\mbox{p}_B$}
\drawline\fermion[\W\REG](21000,12000)[9000]
\drawarrow[\W\ATTIP](\pmidx,\pmidy)
\put(\pmidx,13000){$\mbox{q}_2$}
\drawline\gluon[\W\FLIPPED](12000,12000)[8]
\drawarrow[\W\ATTIP](\pmidx,\pmidy)
\put(\pmidx,13000){$\mbox{q}_1$}
\drawline\gluon[\N\REG](3000,12000)[5]
\drawarrow[\N\ATTIP](\pmidx,\pmidy)
\put(500,\pbacky){$\mbox{p}_{A'}$}
\put(21000,12000){\circle*{1000}}
\put(3000,12000){\circle*{1000}}
\put(12000,12000){\circle*{1000}}
\put(11000,2500){Fig. 4}
\end{picture}

\vspace*{1cm}

\begin{picture}(42000,30000)

\drawline\fermion[\N\REG](3000,24000)[4500]
\drawline\fermion[\NE\REG](3000,24000)[6300]
\drawline\fermion[\S\REG](2800,24000)[4500]
\drawline\fermion[\S\REG](3200,24000)[4500]
\put(3000,24000){\circle*{400}}
\put(4500,18000){a}

\drawline\fermion[\N\REG](15000,19500)[9000]
\drawline\scalar[\E\REG](15000,24000)[3]
\put(15000,24000){\circle*{400}}
\put(18000,18000){b}

\drawline\fermion[\N\REG](34500,21000)[6000]
\drawline\scalar[\E\REG](34500,21000)[3]
\drawline\scalar[\W\REG](34500,21000)[3]
\put(34500,21000){\circle*{400}}
\put(34500,18000){c}

\drawline\fermion[\E\REG](12000,7300)[6000]
\drawline\fermion[\E\REG](12000,7700)[6000]
\drawline\fermion[\NE\REG](12000,7500)[8400]
\drawline\scalar[\W\REG](12000,7500)[3]
\put(12000,7500){\circle*{400}}
\put(12000,4500){d}

\drawline\fermion[\N\REG](37500,4500)[9000]
\put(37500,\pmidy){\circle*{400}}
\drawline\scalar[\W\REG](37500,\pmidy)[3]
\put(33000,4500){e}
\put(19500,1500){Fig. 5}
\put(24000,-7000){36}
\end{picture}

\newpage


\begin{picture}(51000,30000)

\drawline\fermion[\S\REG](3000,21000)[4500]
\drawline\fermion[\N\REG](2800,21000)[3000]
\drawline\fermion[\N\REG](3200,21000)[3000]
\drawline\fermion[\N\REG](3000,24000)[3000]
\drawline\fermion[\NE\REG](3000,24000)[4200]
\drawline\scalar[\E\REG](3000,21000)[2]
\put(3000,21000){\circle*{400}}
\put(3000,24000){\circle*{400}}
\put(3500,16500){p}
\put(6500,27000){k}
\put(5000,19500){q-k}

\put(9000,21000){+}

\drawline\fermion[\N\REG](12000,21000)[6000]
\drawline\fermion[\S\REG](11800,21000)[3000]
\drawline\fermion[\S\REG](12200,21000)[3000]
\drawline\fermion[\S\REG](12000,18000)[1500]
\drawline\scalar[\E\REG](12000,21000)[3]
\put(12000,21000){\circle*{400}}
\put(12000,18000){\circle*{400}}
\drawline\fermion[\NE\REG](12000,18000)[8400]
\put(18000,24500){k}

\put(20000,21000){+}

\drawline\fermion[\N\REG](23000,16500)[10500]
\drawline\fermion[\N\REG](27500,21000)[6000]
\drawline\scalar[\E\REG](23000,21000)[4]
\put(23000,21000){\circle*{400}}
\put(27500,21000){\circle*{400}}
\put(25500,19500){q}
\put(28000,27000){k}

\put(33000,21000){=}

\drawline\fermion[\N\REG](36500,16500)[10500]
\drawline\fermion[\N\REG](41000,21000)[6000]
\drawline\scalar[\E\REG](36500,21000)[4]
\put(36500,21000){\circle*{400}}
\put(41000,21000){\circle*{1000}}
\put(41500,27000){k}

\drawline\scalar[\W\REG](3000,4500)[2]
\drawline\scalar[\E\REG](6000,4500)[2]
\drawline\fermion[\E\REG](3000,4700)[3000]
\drawline\fermion[\E\REG](3000,4300)[3000]
\drawline\fermion[\N\REG](6000,4500)[6000]
\drawline\fermion[\NE\REG](3000,4500)[8400]
\put(3000,4500){\circle*{400}}
\put(6000,4500){\circle*{400}}
\put(6500,10500){k$_1$}
\put(9500,10500){k}

\put(12000,7500){+}

\drawline\scalar[\E\REG](18000,4500)[2]
\drawline\scalar[\W\REG](18000,4500)[2]
\drawline\fermion[\N\REG](17800,4500)[3000]
\drawline\fermion[\N\REG](18200,4500)[3000]
\drawline\fermion[\N\REG](18000,7500)[3000]
\drawline\fermion[\NE\REG](18000,7500)[4200]
\put(18000,4500){\circle*{400}}
\put(18000,7500){\circle*{400}}
\put(18500,10500){k$_1$}
\put(21500,10500){k}

\put(23000,7500){+}

\drawline\fermion[\N\REG](27500,4500)[6000]
\drawline\fermion[\N\REG](32000,4500)[6000]
\drawline\scalar[\W\REG](27500,4500)[1]
\drawline\scalar[\E\REG](27500,4500)[3]
\drawline\scalar[\E\REG](32000,4500)[1]
\put(27500,4500){\circle*{400}}
\put(32000,4500){\circle*{400}}
\put(28000,10500){k$_1$}
\put(32500,10500){k}

\put(35500,7500){=}

\drawline\fermion[\N\REG](40000,4500)[6000]
\drawline\fermion[\N\REG](44500,4500)[6000]
\drawline\scalar[\W\REG](40000,4500)[1]
\drawline\scalar[\E\REG](40000,4500)[3]
\drawline\scalar[\E\REG](44500,4500)[1]
\put(40000,4500){\circle*{400}}
\put(44500,4500){\circle*{1000}}
\put(40500,10500){k$_1$}
\put(45000,10500){k}

\put(24000,500){Fig. 6}
\end{picture}

\vspace*{1cm}

\begin{picture}(42000,21000)
\drawline\fermion[\N\REG](3000,12000)[6000]
\drawline\fermion[\S\REG](3000,12000)[6000]
\drawline\fermion[\N\REG](18000,6000)[12000]
\put(10500,12000){\oval(15000,3000)}
\put(10500,9000){q}
\put(18500,6000){p}
\put(18500,18000){k}
\put(10500,4500){a}

\drawline\fermion[\N\REG](30000,12000)[6000]
\drawline\scalar[\W\REG](30000,12000)[2]
\drawline\fermion[\N\REG](39000,6000)[12000]
\put(34500,12000){\oval(9000,3000)}
\put(34500,9000){q}
\put(39500,6000){p}
\put(39500,18000){k}
\put(34500,4500){b}
\put(21000,1000){Fig. 7}
\put(24000,-6000){37}

\end{picture}

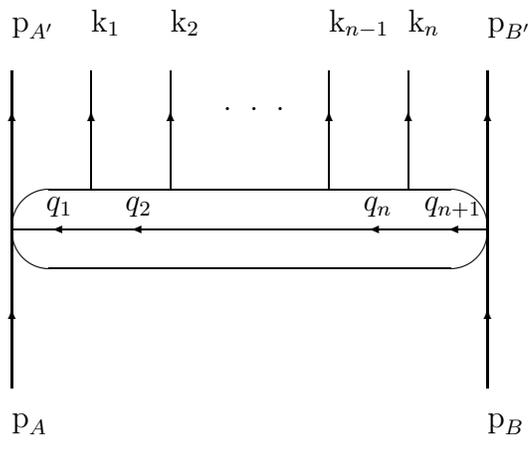
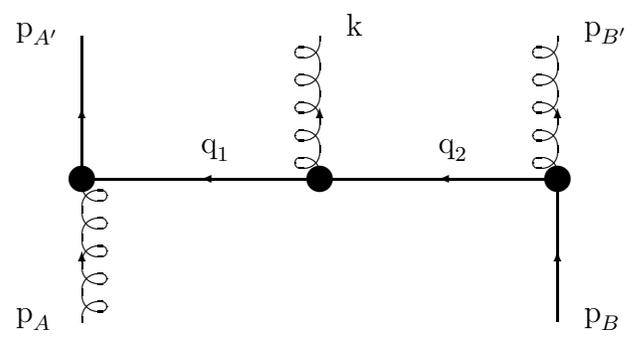
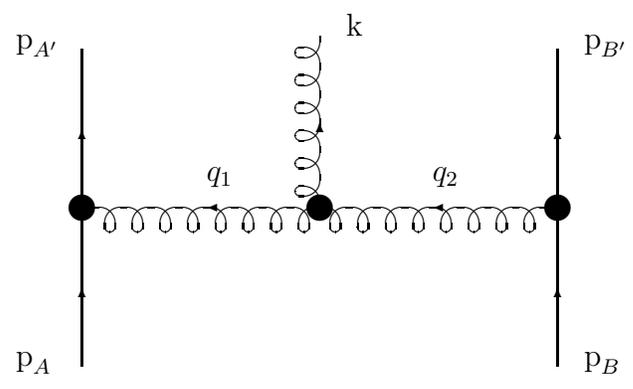
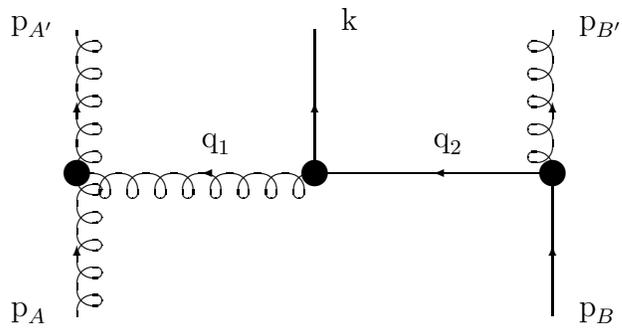
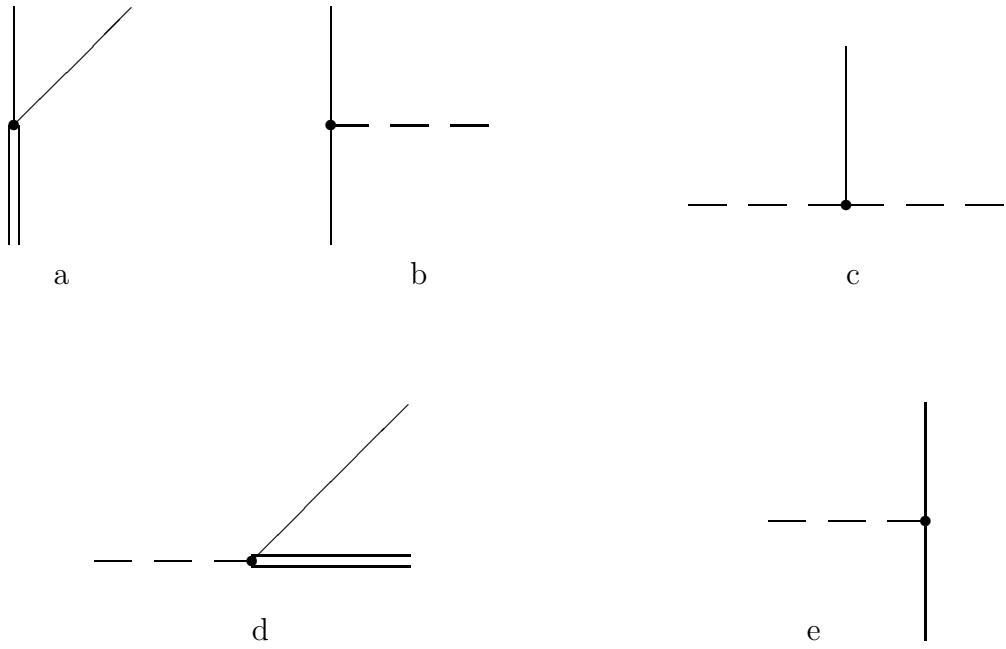
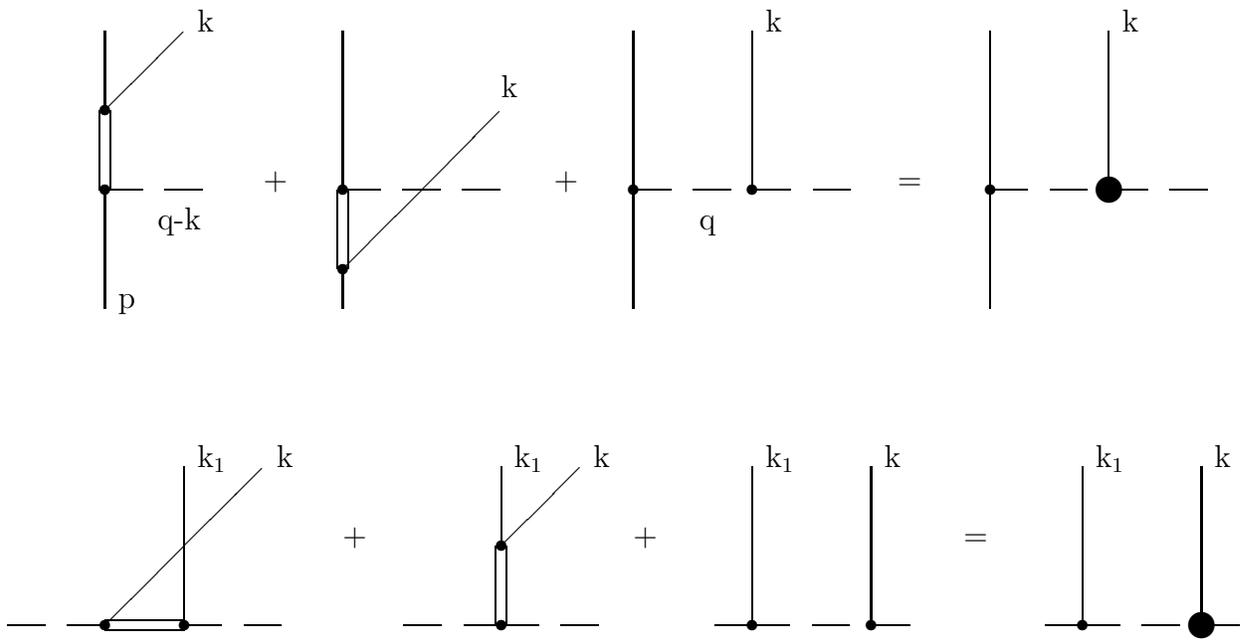
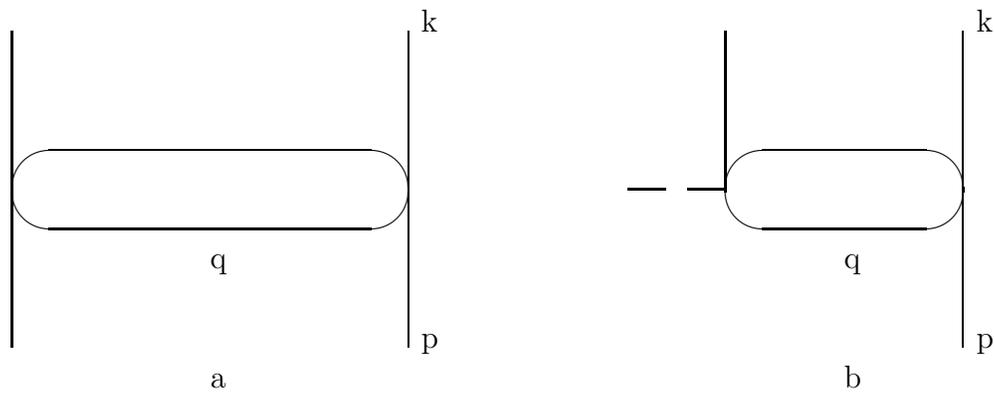
\end{document}